\documentclass[12pt]{article}

\usepackage{amssymb,amscd}
\usepackage{mathrsfs}
\usepackage{amsmath,amsthm}
\usepackage[hypertex]{hyperref}
\usepackage{array,longtable}
\usepackage[dvips]{graphicx}
\usepackage[dvips]{color}
\usepackage{colortbl}
\usepackage{wrapfig}

\newcommand{\version}{November 2, 2006}

\newtheorem{thm}{Theorem}[section]
\newtheorem{lemma}{Lemma}[section]
\newtheorem{cor}{Corollary}[section]
\newtheorem{rem}{Remark}[section]

\newcommand{\Ds}{\mathscr{D}}

\newcommand{\Rh}{\mathbb R}
\newcommand{\Rc}{\mathscr{R}}
\newcommand{\Ch}{\mathbb C}
\newcommand{\Zh}{\mathbb Z}

\newcommand{\Lc}{\mathcal{L}}
\newcommand{\Nc}{\mathcal{N}}

\newcommand{\Hc}{\mathcal{H}}

\newcommand{\Oc}{\mathcal{O}}

\newcommand{\pd}{\partial}

\newcommand{\ch}{\mathop{\mathrm{ch}}\nolimits}

\newcommand{\vol}{\mathop{\mathrm{vol}}\nolimits}

\newcommand{\CSK}{\mathop{\textsl{CSK}\,}\nolimits}

\newcommand{\Lb}{\boldsymbol{\mathcal{L}}}

\renewcommand{\Im}{\mathop{\mathrm{Im}}\nolimits}
\renewcommand{\Re}{\mathop{\mathrm{Re}}\nolimits}

\newcommand{\tr}{\mathop{\mathrm{tr}}\nolimits}

\newcommand{\example}[1]{
\noindent\hspace{-3mm}
\fcolorbox[rgb]{0.8,0.8,0.8}{0.90,0.95,0.85}{
\begin{minipage}[!h]{\textwidth}
#1
\end{minipage}
}
}

\renewcommand{\theequation}{\arabic{section}.\arabic{equation}}

\begin{document}
\title{
\begin{flushright}
{\small Imperial/TP/06/DMB/02}
\\
 \vspace{-6mm}
{\small RUNHETC-2006-28 }
\\
 \vspace{-6mm} {\small hep-th/0611020}
\end{flushright}
\vspace{1cm} \textbf{Type II Actions from 11-Dimensional
Chern-Simons Theories}}
\author{
{Dmitriy M.~Belov$\,{}^a$\; and \;Gregory
W.~Moore$\,{}^b$} \vspace{6mm}
\\
\emph{\normalsize ${}^a$ The Blackett Laboratory, Imperial College London}
\vspace{-2mm}
\\
\emph{\normalsize Prince Consort Road, London SW7 2AZ, UK}
\vspace{5mm}
\\
\emph{\normalsize ${}^b$ Department of Physics, Rutgers University}
\vspace{-2mm}
\\
\emph{\normalsize136 Frelinghuysen Road, Piscataway, NJ 08854, USA}
}


\maketitle
\thispagestyle{empty}

\begin{abstract}
This paper continues the discussion of hep-th/0605038, applying the
holographic formulation of self-dual theory to the Ramond-Ramond
fields of type II supergravity. We formulate the RR partition
function, in the presence of nontrivial H-fields, in terms of the
wavefunction of an 11-dimensional Chern-Simons theory. Using the
methods of hep-th/0605038 we show how to formulate an action
principle for the RR fields of both type IIA and type IIB
supergravity, in the presence of RR current. We find a new
topological restriction on consistent backgrounds of type IIA
supergravity, namely the fourth Wu class must have a lift to
the $H$-twisted cohomology.

\end{abstract}

\vspace{1cm} $~~$ \version

\clearpage

\tableofcontents

\clearpage
\section{Introduction}

Type II supergravity contains differential form fields whose
underlying mathematics has only been properly elucidated in the past
few years. The action principle for these fields includes standard
kinetic terms together with  interesting ``Chern-Simons'' terms.  As
with the analogous expression in M-theory, simply formulating these
terms precisely, in the presence of general background fluxes and
arbitrary topology, is a somewhat subtle problem. While many partial
results exist in the literature, we believe that --- amazingly --- a
complete presentation of the action for the Ramond-Ramond fields of
type II supergravity, even at the 2-derivative level,  has not
appeared previously. The goal of the present paper is to fill this
gap.

The key to formulating the action for type II RR fields is to
understand that the RR field  is a self-dual field quantized by
K-theory, and most naturally formulated in terms of differential
K-theory. This viewpoint has gradually emerged over the past few
years, and is explained in
\cite{Moore:1999gb,Freed:2000tt,Diaconescu:2000wy,
Freed:2000ta,Mathai:2003mu,Freed:2006yc,Freed:2006ya}. The
foundational work of Hopkins and Singer on differential K-theory can
be found in \cite{Hopkins:2002rd}. Pedagogical accounts can be found
in the above references (see, for example, \cite{Freed:2006yc}). In
a previous paper we have shown how Witten's approach to the
formulation of a self-dual field using Chern-Simons theory leads to
a rather simple formulation of an action-principle
\cite{Belov:2006jd}. The present paper applies the methods of
\cite{Belov:2006jd} to the case of RR fields. It follows the
treatment of \cite{Belov:2006jd} very closely: indeed some sections
of the present text were written simply by copy and paste followed
by some small modifications needed to upgrade our previous treatment
to the case of RR fields. We adopted this approach to keep the
paper somewhat self-contained.

One of the key points about a self-dual theory is that there is no
single preferred action. Rather, there is a family of actions,
parameterized --- roughly speaking --- by a choice of Lagrangian
decomposition of fieldspace. The general form of the action we find
is given in Theorem~\ref{thm:91}, equation \eqref{SLD}. Since our notation is
rather heavy let us state this central result somewhat informally.
   The total RR field $G$ is
a trivialization of the RR current $j$:
\begin{equation}
d_H G = j
\end{equation}
where $d_H := d-H$ is the twisted differential, and   $H$ is the
fieldstrength of the NS $B$-field. The difference of two
trivializations of the RR current is  a $d_H$-closed form obeying
some quantization condition. Let us fix a trivialization $G_s$
depending in some definite way on the source $j$.  Then the general
RR field is $G=R + G_s$ where $d_H R=0$. Our choice of ``Lagrangian
decomposition'' constrains $R$ to lie in a certain Lagrangian
subspace (called $V_1$, below) of fieldspace. In addition a choice
of Lagrangian subspace (called $V_2$ below)  also gives a symplectic
decomposition of the total field $G = G^{\rm el} + G^{\rm mag}$.
\footnote{This decomposition should not be confused with the
decomposition $F=B+dt\wedge E$ in Maxwell theory. Here because of
self-duality, $G^{\rm mag}$ are dependent degrees of freedom, and
should be regarded as functions of $G^{\rm elec}$ in a manner
specified by the Lagrangian subspaces $V_1$ and $V_2$.}  Then the
Lorentz-signature action is
\begin{equation}\label{informalaction}
S = -\pi \int G^{\rm el}* G^{\rm el} + \pi \int G^{\rm mag} G^{\rm el}
 - \pi \int G G_s  - \pi \int G_s^{\rm mag} G_s^{\rm el}
\end{equation}
Let us stress that $G$ is \textit{not} self-dual. Rather, it emerges
that when varying the action \eqref{informalaction} the equations of
motion only depend on  the  self-dual projection of it
$\mathcal{F}^+(G)=G^{\text{el}}+*G^{\text{el}}$.  In general it is a
cardinal sin to  include both electric and magnetic degrees of
freedom in a local action. Our formulation avoids that fate because
$R$ is constrained to lie in a Lagrangian subspace.

In the type IIA theory, at large volume and weak coupling, there is
a distinguished subset of Lagrangian decompositions leading to a
form of the IIA action resembling that usually presented in the
literature.  This is the action in \eqref{IIAex1} below. In the type
IIB theory there is no canonical choice of Lagrangian decomposition
for general spacetimes, even at large volume.
This is the reason people
have found it difficult to write an action principle for IIB
supergravity. However, if spacetime has a product structure and we
are working at large distance and weak coupling then there is a
class of natural Lagrangian decompositions, and the corresponding
actions are written in \eqref{SIIBex1} and \eqref{SIIBex2}.

The reliance of the formalism on Lagrangian decompositions is
related to the famous difficulties in formulating a manifestly
Lorentz-invariant action \cite{Marcus:1982yu}. In general,  a choice
of  Lagrangian subspace will break Lorentz symmetry. However a
Lorentz transformation maps one Lagrangian decomposition to another
and since all such decompositions are equivalent we may conclude
that the theory is Lorentz covariant. In this sense, our formalism
is Lorentz covariant in flat space.

One may well ask why one should take so much trouble to write an
action when the equations of motion are well known.
While on-shell quantities suffice for many questions, one
might wish to go off-shell, for example, to compute semiclassical
tunnelling effects. We have in mind as an example computing
transition amplitudes between flux vacua. Moreover, including
higher-order interactions and the computation of quantum corrections
to the leading effects are --- at the very least --- most conveniently
formulated in terms of an action principle.

The remainder of this introduction provides a detailed guide to the
paper.

In section~\ref{sec:RRselfdual} we present a formalism useful for handling the total
self-dual field. For a $10$-dimensional spacetime $X$ we introduce a
space $\Omega(X,\Rc)^j$ of forms which are of even (odd) degree for
$j=0(1) \mod 2$. This is the home of the total $RR$ field. The space
$\Omega(X,\Rc)^j$ is a symplectic vector space with symplectic form
$\omega_j$. Moreover, a metric on spacetime induces a metric on
$\Omega(X,\Rc)^j$   compatible with $\omega_j$, and  defining an
involution (for Lorentzian signature) or a complex structure (for
Euclidean signature) on $\Omega(X,\Rc)^j$. These structures are used
to formulate the self-duality equations. The symplectic geometry of
$\Omega(X,\Rc)^j$ is fundamental to the construction of the action.
The equations of motion of supergravity are succinctly formulated in
terms of a single self-dual fieldstrength $\mathcal{F}^+$.

In section~\ref{sec:RRtop}, which closely parallels \cite{Belov:2006jd}, we
explain the relation of the RR partition function to the
Chern-Simons wavefunction of an $11$-dimensional Chern-Simons
theory. This section restricts attention to topologically trivial
fields in order to explain things in the simplest case. There are
two Chern-Simons theories --- one for type IIA and one for type IIB
supergravity.

In section~\ref{sec:fieldspace} we recall some basic properties of
differential $K$-theory, so that we can extend the discussion of
section~\ref{sec:RRtop} to the topologically nontrivial case. A new
element, compared to \cite{Belov:2006jd} is the possibility of
twisting by a $B$-field. In particular, we will often work with
$H$-twisted cohomology. As above, $H$ is the fieldstrength of the
$B$-field and it is used to form the twisted differential $d_H:
\Omega(X,\Rc)^j \to \Omega(X,\Rc)^{j+1}$. We review some relevant
background material on twisted Chern characters. In section
\ref{sec:defCSAB} we briefly indicate the relation of our discussion
to the work of Hopkins and Singer \cite{Hopkins:2002rd}.

The $11$-dimensional Chern-Simons theory is a ``spin'' Chern-Simons
theory. It requires extra structure to define the Chern-Simons
invariant. That extra structure is a quadratic refinement of a
certain bilinear form. In section \ref{sec:CSACSB} we give a discussion of the
quantization of this ``spin'' Chern-Simons theory, emphasizing the
formulation of the Gauss law for gauge invariant wavefunctions as
the key to the quantization. This leads us, in section~\ref{sec:partfunc}, to
formulate the Chern-Simons wavefunction as a certain theta function
(in infinite dimensions), which can be written as a path integral
over the RR fields. This key result is given in Theorems~\ref{thm:1} and
\ref{thm62}. Sections \ref{sec:CSACSB} and \ref{sec:partfunc} follow closely the analogous treatment in
\cite{Belov:2006jd}. One important improvement we have made is a
better understanding of the basepoint dependence of the theta
functions in the case where the background current is nonzero. See
remark \ref{rem64}.

Section~\ref{sec:Bfield} addresses new complications related to $B$-field
dependence. We show how our formalism allows us to incorporate the
$B$-field equations of motion. This section resolves a problem which
seems to have been overlooked previously, namely, that the famous
one-loop term $\exp[2\pi i \int B X_8]$ of IIA supergravity is not
really well-defined. We show how the ambiguity in its definition is
cancelled by a compensating ambiguity in the RR partition
function. Moreover, this improved understanding leads to a new
topological consistency condition for IIA string theory. That
consistency condition states that the fourth Wu class $\nu_4$ must
have a lift to $H$-twisted cohomology of the form $(\lambda + 2\rho)+\dots$
where the ellipsis denotes higher degree terms and $\rho$ denotes a closed
$4$-form with integral periods. In
other words, we claim that consistent string backgrounds must
satisfy
\begin{equation}
[H\wedge (\lambda+2\rho)]_{\text{de Rham}} = 0
\end{equation}
together with some conditions involving Massey products.

In section~\ref{sec:action} we combine our formalism for RR fields with the
general discussion of \cite{Belov:2006jd} to write an extremely
compact formula for the full RR action:
\begin{equation}\label{action}
S(R) = \pi \omega_j(R,\mathcal{F}^+(R)).
\end{equation}
Here $\omega_j$ is the symplectic form on $\Omega(X,\Rc)^j$ and $R$
is the total RR field. It is the fieldstrength of a differential
$K$-theory class, and hence its ``periods'' (more precisely, its
$H$-twisted cohomology class) is quantized by twisted $K$-theory.
$\mathcal{F}^+(R)$ is the self-dual projection of $R$ \eqref{SMaction}. Its
``periods'' depend on the metric.

The action \eqref{action} depends on two important choices. First,
one must choose a Lagrangian subspace $V_2 \subset \Omega(X,\Rc)^j$.
This projects to a Lagrangian lattice $\Gamma_2$ in the space of
$H$-twisted harmonic forms. The second choice is a Lagrangian
subspace $\Gamma_1$ in the space of $H$-twisted harmonic forms
complementary to $\Gamma_2$. Such a subspace canonically determines
a Lagrangian space $V_1$ of $d_H$-closed forms. A crucial part of
our formulation of the RR field is that $R$ must take values in
$V_1$. One pitfall should be avoided: $V_1$ and $V_2$ do not form a
Lagrangian decomposition of $\Omega(X,\Rc)^j$: Their sum is not the
whole space and they have a nontrivial intersection $V_1\cap V_2
\not= \{ 0 \}$.

Our action is easily extended to include the presence of RR
current, including a background current $\check \mu$ induced by the
topology of spacetime (and the choice of Lagrangian spaces
$V_1,V_2$). The action in the presence of RR currents is given by
Theorem 9.4. The shift $R \to R-\sigma_\bullet$ that appears in that
theorem accounts for the $\frac12$-integer shifts in the
quantization of RR fields.

In section~\ref{sec:examples} we spell out our action for IIA and IIB supergravity
using some natural Lagrangian decompositions. In order to stress the
point that the action depends on the choice of Lagrangian subspaces
of fieldspace we illustrate two natural decompositions on product
spacetimes. We hope that these expressions will be understandable
even to people who do not have the patience to master our formalism.
Section~\ref{sec:metric} on the metric dependence and stress energy tensor closely
follows the discussion in \cite{Belov:2006jd}. Section~\ref{sec:conc} concludes
by mentioning some directions for further research.

Appendix A records some nontrivial examples of twisted cohomology on
spaces related to twisted tori. In appendix B we comment on the
one-loop determinants for the RR field. These follow, in our
treatment, from normalizing the Chern-Simons wavefunction. Finding
an expression in terms of determinants is considerably more involved
than in the previous case \cite{Belov:2006jd} because the
differential $d_H$ has inhomogeneous degree.

To conclude, we comment on some related literature. Some aspects of
our discussion resemble the ``democratic formulation'' of
\cite{Bergshoeff:2001pv}. What our treatment adds is a careful
discussion of how higher degree forms depend on lower ones (in the
relevant Lagrangian decomposition) and a framework in which
topologically nontrivial fieldstrengths can be included. Our
discussion also has overlap with previous discussions of the twisted
$K$-theory parition function \cite{Moore:2002cp,Mathai:2003mu} but
these previous discussions were incomplete. Our formalism also
nicely resolves some puzzles raised by de Alwis
\cite{deAlwis:2006cb}. (de Alwis has also arrived at a resolution of
the same difficulties within his own formalism.)

\clearpage
\section{Self-duality constraints on RR fieldstrengths}
\label{sec:RRselfdual}\setcounter{equation}{0}

In this section we introduce some formalism which will be used
throughout the paper which is useful for handling the full set of RR
fields and clarifies the self-duality of the total RR field.  We
will show that the entire set of  supergravity equations depends on
the RR field only through its self-dual combination.

\paragraph{Symplectic structure.}

Let $X$ be a  $10$-dimensional oriented manifold. We will be working
with differential forms of even or odd degree on $X$. To keep track
of the differential form degree it is convenient to introduce a
formal variable $u$ of degree $2$. We will write even and odd forms
as a series in $u^{-1}$:
\begin{equation}
w=w_0+u^{-1}w_2+\dots+u^{-5}w_{10}
\quad\text{or}\quad w=u^{-1}w_1+u^{-2}w_3+\dots+u^{-5}w_9
\end{equation}
where $w_p$ is a $p$-form. The total degree of $w$ (degree of
differential form + degree of $u$)  is $0$ for the first series and
$-1$ for the second. So we introduce a graded ring
$\Rc=\Rh[u,u^{-1}]$ and denote by $\Omega(X;\Rc)^j$ the space of all
differential forms of \textit{total} degree $j$. In
section~\ref{sec:fieldspace} we will see that the graded ring $\Rc$
naturally appears in  differential $K$-theory ($u^{-1}$ is the
generator of $K^{-2}(pt)=\Zh[u^{-1}]$ and is called the Bott
element). The space of forms $\Omega(X;\Rc)^j$ of \textit{total}
degree $j$ is a symplectic vector space with   symplectic form
\begin{equation}
\omega_j =\frac{1}{2}\int_X \delta w\wedge \phi_j(\delta w)
\label{omegaj}
\end{equation}
where for a $10$-manifold $X$ $\phi_j$ is a map
$\phi_j:\Omega(X;\Rc)^j\to \Omega(X;\Rc)^{10-2j}$ and
\begin{equation}
\phi_j(w):=(-1)^{j(j-1)/2}u^{5-j}(w|_{u\to-u}).
\label{phijdef}
\end{equation}
The integral is assumed to pick up the coefficient of $u^0$. The
factor $(-1)^{j(j-1)/2}$ is included to ensure that the symplectic
form $\omega_j$ has the property $\omega_j(d
\lambda,w)=\omega_{j+1}(u\lambda,dw)$. For example,
\\
\example{
\begin{subequations}
\begin{align}
\omega_0(v,w)&=\int_X(-v_0 w_{10}+v_2 w_8-
v_4 w_6+v_6 w_4-v_8 w_2 +v_{10} w_0);
\\
\omega_{-1}(v,w)&=\int_X(v_1 w_{9}
-v_3 w_7+v_5 w_5 -v_7 w_3+v_{9} w_1).
\end{align}
\end{subequations}
}

\noindent Note the signs. They guarantee that the symplectic form $\omega_j$
is invariant under the $b$-transform:
\begin{equation}
\omega_j(e^{u^{-1}b}v,e^{u^{-1}b}w)
=\omega_j(v,w)
\label{btrinv}
\end{equation}
where $b\in\Omega^2(X)$. So the $b$-transform is a symplectomorphism.

\paragraph{Self-duality of the Ramond-Ramond field.}

We denote by $(M,g)$ a  $10$-dimensional oriented Lorentzian
manifold. A Lorentzian metric $g$ on $M$ together with the string
scale $\ell_s$ define an indefinite metric $g_j$ on the vector space
$\Omega(M;\Rc)^j$:
\begin{equation}
g_j(v,w):=\int_X v\wedge \check{\imath}(w)
\quad\text{where}\quad
\check{\imath}(w):= (*_{\ell_s^{-2}g} w)|_{u\mapsto u^{-1}}.
\label{metricL}
\end{equation}
Here $\check{\imath}:\Omega(X;\Rc)^j \to\Omega(X;\Rc)^{10-j}$ is the
Hodge star operation for the rescaled metric $\ell_s^{-2}g$ followed
by the substitution $u\to u^{-1}$. In our conventions all forms are
dimensionless. Therefore the Hodge star operator for $g$ is a
dimensionfull operation: the dimension of $* w_p$ is
$(\text{length})^{10-2p}$. But the Hodge star operator for the
rescaled metric $\ell_s^{-2}g$ is dimensionless --- that's why it
is used in \eqref{metricL}. In   components the metric
\eqref{metricL} has the following very simple form

\example{
\begin{subequations}
\begin{align}
g_0(v,w)&=\int_X \bigl[
\ell_s^{-10}\,v_0\wedge * w_0+\ell_s^{-6}\, v_2\wedge * w_2+
\dots + \ell_s^{10}\,v_{10}\wedge * w_{10}
\bigr];
\\
g_{-1}(v,w)&=\int_X \bigl[
\ell_s^{-8}\,v_1\wedge * w_1+\ell_s^{-4}\, v_3\wedge * w_3+
\dots + \ell_s^{8}\,v_{9}\wedge * w_{9}
\bigr].
\end{align}
\label{Metdef}
\end{subequations}
}

The indefinite metric together with the symplectic form define an
involution $I:\Omega(M;\Rc)^j \to\Omega(M;\Rc)^j$ by
\begin{equation}
g_j(v,w)=\omega_j(I(v),w)\quad\Rightarrow\quad
I(v):=-(\phi^{-1}_j\circ\check{\imath})(v).
\label{Idef}
\end{equation}
\example{
For $j=0$ and $j=-1$ the involution is
\begin{equation}
I_0(w)=\sum_{p=0}^5u^{-p}(-1)^{p+1}\,\ell_s^{10-4p}*w_{10-2p}
\quad\text{and}\quad
I_{-1}(w)=\sum_{p=1}^5u^{-p}(-1)^{p}\,\ell_s^{12-4p}*w_{11-2p}.
\label{Iexml}
\end{equation}
} Using this involution we can decompose the vector space
$\Omega(M;\Rc)^j$ into a sum of eigenspaces corresponding to $I=+1$
and $I=-1$. We will refer to  the forms $\{\mathcal{F}^+\}$ which
satisfy the equation $I(\mathcal{F}^+)=\mathcal{F}^+$ as to the
\textit{self-dual} forms \footnote{Another way of defining an
involution $I$ and self-duality condition
$I\mathcal{F}^+=\mathcal{F}^+$ is as follows.  The volume element on
a Lorentzian $(4k+2)$-dimensional manifold  defines the involution
in the Clifford algebra $\bar{\Gamma}=c(\vol(g))$, where $c(\omega)$
denotes Clifford multiplication by a form $\omega$.  Now for
$w\in\Omega(X;\Rc)^j$ we define involution $I$ by
\begin{equation*}
c(Iw):=\bar{\Gamma}c(w).
\end{equation*}
Using the relation $\bar{\Gamma}c(w_p)=(-1)^{p(p+1)/2}c(*w_p)$ one
can show that the involution defined by $\bar{\Gamma}$ yields
exactly $I_0$ and $I_{-1}$ written in \eqref{Iexml}. }.

The main observation is that \textit{all} the  equations of type II
supergravity can be rewritten using only the self-dual form
$\mathcal{F}^+$:
\begin{subequations}
\begin{align}
\text{Bianchi id + RR eom: }&\quad (d-u^{-1}H)\mathcal{F}^+=0;
\\
\text{eom for $H$-flux: }&\quad
\ell_s^{-4}d(e^{-2\phi}*H)=-\frac12\,\Bigl[
\mathcal{F}^+\wedge\phi_j(\mathcal{F}^+)\Bigr]_{\text{coef. of }u^1}
+\text{Gravitational  correction}; \label{Heom}
\\
\text{stress-energy tensor: }&\quad
\delta g^{\mu\nu}T_{\mu\nu}\,\vol(g)
=\Bigl[\xi_g\mathcal{F}^+\wedge \check{\imath}(\mathcal{F}^+)
\Bigr]_{\text{coef. of }u^0}
\end{align}
\end{subequations}
where $\xi_g=(\delta g^{-1}g)^{\mu}{}_{\nu}\,dx^{\nu}\wedge i(\frac{\pd}{\pd x^{\mu}})$
(for the properties of this operator see section~\ref{sec:metric}).
The following examples illustrate this observation:
\vspace{3mm}
\\
\example{
For type IIA the self-dual field is
\begin{equation}
\mathcal{F}^+=R_0+u^{-1}R_2+u^{-2}R_4
+\ell_s^{-2}  u^{-3}\,*R_4
-\ell_s^{-6}  u^{-4}\,*R_2
+\ell_s^{-10}u^{-5}\,*R_0.
\label{F+IIA}
\end{equation}
The equation $d_H\mathcal{F}^+=0$ yields Bianchi identities
together with equations of motion
\begin{equation*}
dR_0=0,\quad dR_2=HR_0,\quad dR_4=HR_2,\quad \ell_s^{-2}\,d*R_4=HR_4,
\quad \ell_s^{-4}d*R_2=-H\wedge *R_2.
\end{equation*}
The equation  \eqref{Heom} is
\begin{equation*}
\ell_s^{-4}d(e^{-2\phi}*H)
=\ell_s^{-6}R_0\wedge *R_2+
\ell_s^{-2}R_2\wedge *R_4-\frac12\,R_4\wedge R_4
+X_8(g)
\end{equation*}
where $X_8(g)=\frac{1}{48}(p_2-\lambda^2)$ and
$\lambda=-\frac{p_1}{2}$ and $p_i$ are differential forms
representing the Pontrjagin class constructed from traces of the
curvature.  The stress-energy tensor is given in \eqref{TIIA}. }
\example{
For type IIB the self-dual field is
\begin{equation}
\mathcal{F}^+=u^{-1}R_1+u^{-2}R_3+u^{-3}\mathcal{F}_5^+
+\ell_s^{-4}  u^{-4}\,*R_3
-\ell_s^{-8}  u^{-5}\,*R_1
\label{F+IIB}
\end{equation}\
where $* \mathcal{F}_5^+=-\mathcal{F}_5^+$.
The equation $d_H\mathcal{F}^+=0$ yields
Bianchi identities
together with equations of motion
\begin{equation*}
dR_1=0,\quad dR_3=HR_1,\quad d\mathcal{F}_5^+=HR_3,\quad
\ell_s^{-4}\,d*R_3=H\mathcal{F}_5^+,
\quad \ell_s^{-4}d*R_1=-H\wedge *R_3.
\end{equation*}
The equation  \eqref{Heom} is
\begin{equation*}
\ell_s^{-4}d(e^{-2\phi}*H)
=\ell_s^{-4}R_1\wedge *R_3-
R_3\wedge \mathcal{F}^+_5.
\end{equation*}
The stress-energy tensor is given in \eqref{TIIB}.
}

Note that \eqref{F+IIA} and \eqref{F+IIB}
are just particular parameterizations of the self-dual field:
the self-dual field can always be written in the form
$\mathcal{F}^+=F+IF$ for $F\in\Omega(M;\Rc)^j$
--- however $F$ is not unique.

\paragraph{Complex structure.}

Our quantization of   Chern-Simons will be in Euclidean signature.
So in the next five sections we will be working on a compact
Riemannian manifold $(X,g_E)$. Similarly to \eqref{metricL}, the
Riemannian metric $g_E$ on $X$ define a Riemannian metric $g_j$ on
the vector space $\Omega(X;\Rc)^j$.

A   complex structure  $J:\Omega(X;\Rc)^j \to\Omega(X;\Rc)^j$
compatible with the metric and symplectic form is defined by
\begin{equation}
g_j(v,w)=\omega_j(J(v),w)\quad\Rightarrow\quad
J(v):=-(\phi^{-1}_j\circ\check{\imath})(v).
\label{Jdef}
\end{equation}
In components $J$ is given by the expression similar to \eqref{Iexml}.

\clearpage
\section{Formulating the partition function for topologically trivial fields}
\label{sec:RRtop}\setcounter{equation}{0}

Before describing the general case, we first explain our strategy in
the case where all the RR fields are topologically trivial.

In \cite{Witten:1996hc} Witten argued that the partition function
for a self-dual field on a Riemannian $(4k+2)$-dimensional manifold,
 as a function of external currents,
can be obtained from a certain abelian Chern-Simons theory in one
dimension higher (see also section~2 of \cite{Belov:2006jd} for more
details). The main result of this section is that the partition
function of Ramond-Ramond fields of type IIA/IIB supergravity can be
obtained from a certain abelian Chern-Simons theory in
$11$-dimensions. There are two natural Chern-Simons theories in $11$
dimensions: one of them --- CSA --- yields the partition function
for type IIA,  and another  --- CSB --- for type IIB. These
Chern-Simons theories depend on different gauge fields:
\begin{center}
\renewcommand{\arraystretch}{1.1}
\begin{tabular}{ccc}
\hline \rowcolor[gray]{0.85} & CSA & CSB
\\
\hline gauge field & $A\in\Omega^{\textsl{ev}}(X)$ &
$A\in\Omega^{\textsl{odd}}(X)$
\\
\hline
\end{tabular}
\end{center}
For topologically trivial gauge fields the Chern-Simons functional
is given by
\begin{equation}
e^{2\pi i\CSK^{j+1}(A)}=\exp\left[2\pi i\, \frac12\int_Y u A\wedge
\phi_{j+1}(d_HA)\right] \label{CStoptriv}
\end{equation}
where $j=0$ for IIA and $j=-1$ for IIB, and $d_H = d- u^{-1} H$ is
the twisted differential. (The NS $H$-field is extended from $X$ to
$Y$.)

In this section we start from the Euclidean action principal in
which forms of \textit{all} degrees appear (even degree for type IIA
and odd degree for type IIB). This is known as a ``democratic''
formulation of supergravity \cite{Bergshoeff:2001pv}. Next we couple
the RR fields to the external RR current in such a way that only the
imaginary anti self-dual part of the Ramond-Ramond form is coupled
to the current. We will then show that the partition function as a
function of this external current is a section of the Chern-Simons
line bundle.

\paragraph{``Democratic'' formulation.}
The space of Ramond-Ramond fields is fibered over the space of
$B$-fields. We will describe this structure in detail in
sections~\ref{sec:fieldspace} and \ref{sec:Bfield}. In this
paragraph we only need $H\in\Omega^3_{\Zh}(X)$
 --- the curvature of the $B$-field.
Topologically trivial Ramond-Ramond fields are  described by a gauge potential
$C\in\Omega(X;\Rc)^{j-1}$
with curvature $d_HC\in \Omega(X;\Rc)^j_{d_H}$ where $d_H=d-u^{-1}H$.
Following \cite{Witten:1996hc} we start from the following action functional for
the $C$-field\footnote{This functional
is the bosonic part of pseudo-action \cite{Bergshoeff:2001pv},
known also as ``democratic'' formulation of supergravity.}
\begin{equation*}
S(C)=\frac{\pi}{2}\int_X d_HC\wedge \check{\imath}(d_HC)
\end{equation*}
where $\check{\imath}$ is defined in \eqref{metricL}. We now
introduce  a topologically trivial $U(1)$ external gauge field
$A\in\Omega(X;\Rc)^{j}$ with gauge transformation law $\delta
A=d_H\lambda$ and $\delta C=\lambda$ for
$\lambda\in\Omega(X;\Rc)^{j-1}$, so that the field $C$ has charge
one under this $U(1)$. Consider the Lagrangian
\begin{equation}
e^{-S(C,A)}=\exp\Bigl[
-\frac{\pi}{2}
\,g_j(d_H C-A,\,d_HC-A)
-i\pi\omega_j(d_HC,A)
\Bigr]
\label{lagr}
\end{equation}
where $g_j$ and $\omega_j$ are the Riemannian metric \eqref{metricL}
and symplectic form \eqref{omegaj}. To understand the effect of the
topological interaction it is useful to rewrite this action in
complex coordinates. Using the complex structure $J$ defined in
\eqref{Jdef} we can rewrite $R=d_HC$ as $R=R^++R^-$ and $A=A^++A^-$
where
\begin{equation}
R^{\pm}=\frac12(R\mp iJ\cdot R)
\end{equation}
and similarly for $A^{\pm}$. Here $A$ and $C$ are real forms and
thus $(A^-)^*=A^{+}$ and $(R^-)^*=R^+$. In this notation we obtain
\begin{equation}
e^{-S(C,A^+,A^-)}:=\exp\Bigl[
i\pi\omega_j(R^-,R^+)
+i\pi \omega_j(A^-,A^+)
+2\pi i\omega_j(A^+,R^-)
\Bigr].
\label{Sca}
\end{equation}
Once we obtain this expression we can treat $A^+$ and $A^-$
as \textit{independent} complex variables while keeping $R$ real.

It follows that the holomorphic dependence of the partition function
\begin{equation}
\mathcal{Z}(A^+,A^-)=\int_{\text{top. trivial.}} \Ds C\,e^{-S(C,A^+,A^-)}
\label{ZA}
\end{equation}
on $A^+$ represents the coupling to the anti-self-dual degree of
freedom. The covariant derivatives are
\begin{equation}
D^-=\delta^--i\pi\omega_j(\delta A^-,A^+)
\quad\text{and}\quad
D^+=\delta^+-i\pi\omega_j(\delta A^+,A^-)
\label{connectionA}
\end{equation}
where $\delta^{\pm}$ denotes the usual differential with respect to
$A^{\pm}$. The partition function obeys the holomorphic equation
\begin{equation}
D^{-}\mathcal{Z}(A)=0.
\end{equation}
This easily follows since the action \eqref{Sca}
satisfies this equation.
Since $[D^-,D^-]=0$ the connection \eqref{connectionA}
defines a holomorphic
line bundle $\mathcal{L}$ over the space of \textit{complexified} gauge fields $\{A^+\}$.

The partition function is a holomorphic section of $\mathcal{L}$.
The fact that the partition function is not a function but a section
of a line bundle is related to the fact that the action \eqref{lagr}
is not gauge invariant. If $X$ is a closed manifold then under small
gauge transformation $\delta C=\lambda$, $\delta A=d_H\lambda$ it
transforms as
\begin{equation}
\delta S=i\pi \int_X u\lambda\wedge
\phi_{j+1}(d_HA)\quad\Rightarrow\quad
\mathcal{Z}(A+d_H\lambda)=\mathcal{Z}(A)\,e^{-i\pi
\omega_{j+1}(u\lambda,d_HA)}.
\end{equation}
Thus the partition function
on the real slice $A^-=(A^+)^*$
obeys the non standard gauge-invariance:
\begin{equation}
\bigl[D_{d_H\lambda}+2\pi i\, \omega_{j+1}(u\lambda,d_HA)\bigr]\mathcal{Z}(A)=0
\label{gaugeinv}
\end{equation}
where $D=\delta-i\pi \omega_j(\delta A,A)$
with $\delta$ being the differential on the space of gauge fields,
and $D_{d_{H}\lambda}$ denotes the covariant derivative $D$ evaluated on the vector field
$d_H\lambda$.
The connection $D$ has a nonzero curvature
\begin{equation}
D^2=2\pi i \, \omega_j.
\label{omega1}
\end{equation}

Equation \eqref{gaugeinv} represents the
quantum equation of motion for the self-dual field.
Indeed, substituting \eqref{Sca} and $D=D^++D^-$ one finds
\begin{equation*}
d_H\langle R^--A^-\rangle+ d_HA=0
\end{equation*}
where $\langle\dots \rangle$ denotes a normalized correlation function
for the functional integral \eqref{ZA}.

\paragraph{Complexification of the gauge group.}
The fact that the partition function is a \textit{holomorphic}
section of $\mathcal{L}$ allows us to complexify the
gauge group. Recall that originally the partition function
$\mathcal{Z}(A)$ was a function of a real gauge field $A$. By
writing $A=A^++A^-$ we realized that it depends holomorphically on
the complex field $A^+$. This   means that $A^+$ and $A^-$ can be
considered as independent complex variables so $(A^-)^* \not= A^+$.
This in turn allows us to complexify the gauge group. Originally,
the gauge transformations were given by a real form
$\lambda\in\Omega(X;\mathcal{R})^{j-1}$: $A\mapsto A+d_H\lambda$.
Complexification of the
gauge group means that now we have two complex gauge parameters
$\lambda^+$ and $\lambda^-$, and gauge transformations
\begin{equation*}
A^+\mapsto A^++\frac12(d_H\lambda^+-iJ\cdot (d_H\lambda^+))\quad
\text{and}\quad
A^-\mapsto A^-+\frac12(d_H\lambda^-+iJ\cdot (d_H\lambda^-)).
\end{equation*}
Notice that the field strength $F=d_HA^++d_HA^-$ is not invariant under
the complex gauge transformation:
\begin{equation*}
F\mapsto F+ \frac{1}{2 i}\,d_H [J(d_H\lambda^+-d_H\lambda^-)].
\end{equation*}
Evidently, by a complex gauge transformation we can transform a
topologically trivial gauge field $A$ to be flat: $d_HA=0$.

To proceed further we need to modify the partition function
\eqref{ZA} to include a sum over topological sectors. This step is
quite  nontrivial, and requires conceptual changes.  We postpone the
details of the construction to the next sections. The partition
function takes the schematic form
\begin{equation}
\mathcal{Z}(A):=\sum_{a\in K^{j,h}(X)} \Omega(a)\,
\int_{\text{fixed top. sector}}\Ds C_a \,e^{-S(R,A)} \label{p.func}
\end{equation}
where $\Omega:K^{j,h}(X)\to \{\pm 1\}$ is a crucial phase factor
discussed in detail in section~\ref{sec:CSACSB}.

\paragraph{The partition function as a holomorphic section of a line bundle.}
The space of   topologically trivial flat gauge fields is a torus:
\begin{equation}
\mathcal{W}^{j}(X)
=\Omega(X;\mathcal{R})^j_{d_H}/\Omega(X;\mathcal{R})^j_{d_H,\Zh}
\end{equation}
which is a quotient of the space of  $d_H$-closed forms of   total
degree $j$, $\Omega(X;\mathcal{R})^j_{d_H}$, by the group of  large
gauge transformations $A\mapsto A+R$ where $R$ is a $d_H$-closed
form of total degree $j$ with a certain  quantization
condition \footnote{For the definition of a $d_H$-closed form with
``integral periods'' see section~\ref{sec:fieldspace}. The facts
quoted above follow from the perfect pairing on twisted $K$-theory.}. The
symplectic form $\omega_j$ takes integral values on
$\Omega(X;\mathcal{R})^j_{d_H,\Zh}$. Thus the partition function is
a holomorphic section of the line bundle $\Lb$ over the complex
torus $\mathcal{W}^{j}_{\Ch}(X)$ which is obtained from the real
torus $\mathcal{W}^{j}(X)$ by using the complex structure $J$. Note
that $\dim _{\Rh} \mathcal{W}^j_{\Ch}$ is not given by Betti numbers
but depends on the de Rham cohomology class $[H]\in H^3_{dR}(X)$.
The line bundle $\Lb\to \mathcal{W}^{j}(X)$ has a nonzero first
Chern class $c_1(\Lb)=[\omega_j]_{dR}$. The symplectic form
$\omega_j$ is of type $(1,1)$ in the complex structure $J$. From the
Kodaira vanishing theorem and the index of $\bar{\pd}$-operator it
follows that
\begin{equation}
\dim H^0(\mathcal{W}_{\Ch}^{j},\mathcal{\Lb})=
\int_{\mathcal{W}^{j}}e^{c_1(\mathcal{\Lb})}\mathrm{td}(T\mathcal{W}^{j})
=\int_{\mathcal{W}^{j}}e^{\omega_j}=1.
\label{dimL}
\end{equation}

Following the ideas of \cite{Witten:1996hc} we expect that this
construction describes the partition function of an imaginary anti
self-dual RR field. {}From \eqref{dimL} it follows that the line
bundle $\Lb$ has a \textit{unique} holomorphic section. This
holomorphic section is the partition function for  RR fields
as a function of the RR current.

Therefore to construct a partition function for RR fields
we need to
\begin{enumerate}
\item construct a line bundle $\Lb$ over the torus $\mathcal{W}^{j}(X)$ equipped
with norm and hermitian connection $D$ with curvature $-2\pi i \omega_j$.

\item choose the complex structure \eqref{Jdef} on the torus $\mathcal{W}^{j}(X)$.
Using the connection $D^-$ we can define holomorphic
sections of $\Lb$.
\end{enumerate}

A natural geometrical way of constructing the line bundle and
connection on it is to use Chern-Simons theory in one higher
dimension as we describe next.

\paragraph{Relation to Chern-Simons theory.}
A lot of information about
the line bundle $\mathcal{L}\to\Omega(X;\Rc)^j$ is encoded in the topological term
\begin{equation}
e^{-i\pi \omega_j(d_HC,A)}.
\label{topterm}
\end{equation}
Recall that this exponential is not gauge invariant: under the gauge transformation
$\delta A=d_H\lambda$ and $\delta C=\lambda$ it is multiplied by
\begin{equation}
e^{-i\pi \omega_{j+1}(u\lambda,d_HA)}.
\label{gtr}
\end{equation}
This extra phase coming from the gauge transformation looks like the
boundary term of a level $1$ abelian Chern-Simons theory
 in one dimension higher. Indeed, let $Y$ be an $11$-manifold
with boundary $X$. Consider the following topological action for a
topologically trivial gauge field $A\in\Omega(Y;\Rc)^j$
\begin{equation}
e^{2\pi i \CSK^j_Y(A)}:=
\exp\left[
2\pi i\,\frac12\int_Y uA\wedge \phi_{j+1}(d_H A)\right]
\end{equation}
This exponential is not gauge invariant on a manifold with boundary.
Under the gauge transformation $A\mapsto A+d_H\lambda$ it multiplies  by the
inverse of boundary term \eqref{gtr}. The Chern-Simons functional on a manifold
with boundary is most naturally considered as a section of
the Chern-Simons line bundle over the space of gauge fields
$\{A_X\}$ on the boundary $X$. Our simple calculation shows that
the Chern-Simons line bundle is isomorphic to $\mathcal{L}^{-1}$.

The above discussion can be extended to include the coupling to RR
currents (in particular, the coupling to D-branes). We give this in
section~\ref{sec:defCSAB} below for the topologically nontrivial case.

\section{Fieldspace and gauge transformations}
\label{sec:fieldspace}\setcounter{equation}{0}
To proceed further we need to generalize the above construction to
allow topologically nontrivial gauge fields $C$ and $A$.
In the previous section we saw that the gauge field $A$ plays the role
of an external current for the RR field.
The space of gauge fields $A$ is fibered over the space of $B$-fields.
Recall that space of gauge equivalence classes of $B$-fields is
an infinite dimensional abelian group $\check{H}^3(Y)$.
In this paper we will denote by $\check{B}\in\check{Z}^2(Y)$ a representative of
the differential cohomology class $[\check{B}]\in \check{H}^3(Y)$,
by $h\in H^3(Y;\Zh)$ its characteristic class and by $H\in\Omega^3_{\Zh}(Y)$
its curvature. The space $\check{Z}^2(Y)$ is the space of differential cocycles.
Its definition depends very much on a choice of a model for the differential cohomology
(see \cite{Hopkins:2002rd} and section~3 in \cite{Belov:2006jd} for more details).

\paragraph{Differential K-group.}
The set of gauge-inequivalent fields (or gauge inequivalent currents)
is an infinite dimensional abelian group
$\check{K}^{j+1,\check{B}}(Y)$, known as a twisted differential
$K$-group. For a  pedagogical introduction
to differential K-theory  see section~5 of \cite{Freed:2006yc}. This group
can be described by two exact sequences:
\begin{itemize}
\item Field strength exact sequence
\begin{equation}
0\to \mathop{\underbrace{K^{j,h(\check{B})}(Y;\Rh/\Zh)}}_{\text{flat
fields}}\to\check{K}^{j+1,\check{B}}(Y) \stackrel{F}{\longrightarrow}
\Omega(Y;\Rc)^{j+1}_{d_H,\Zh} \to 0. \label{fieldstr}
\end{equation}
Here $\Rc=\Rh[u,u^{-1}]$ where $u$ is the inverse Bott element of degree $2$.
Every twisted differential $K$-character $[\check{A}]$ has a field strength
$F([\check{A}])$ which is a $d_H$-closed form of total degree $j+1$ with some
quantization condition. We will explain this quantization condition
below.

\item Characteristic class exact sequence
\begin{equation}
0\to
\mathop{\underbrace{\Omega(Y;\Rc)^j/\Omega(Y;\Rc)^j_{d_H,\Zh}}}_{
\text{topologically trivial}} \to
\check{K}^{j+1,\check{B}}(Y)\mathop{\longrightarrow}^{x}
K^{j+1,h(\check{B})}(Y)\to 0 \label{charclass}
\end{equation}
Every twisted differential $K$-character $[\check{A}]$ has a characteristic class
$x([\check{A}])$ which is an element of the twisted $K$-group
$K^{j+1,h(\check{B})}(Y)$.
\end{itemize}
The field strength and characteristic class are compatible in the
sense that  the twisted Chern character $\ch_{\check{B}}(x)$
 must coincide with the twisted cohomology class
$[F]_{d_H}$ defined by the field strength:
\begin{equation*}
\sqrt{\hat{A}}\ch_{\check{B}}(x)=[F]_{d_H}.
\end{equation*}
Here $\ch_{\check{B}}:K^{j+1,h(\check{B})}(Y)\to
H(Y;\Rc)^{j+1}_{d_H}$ is the  twisted Chern character,  which we
define below. Putting together the two sequences we can visualize
the infinite dimensional abelian group
$\check{K}^{j+1,\check{B}}(Y)$ as
\begin{figure}[!!h]
\centering
\includegraphics[width=400pt]{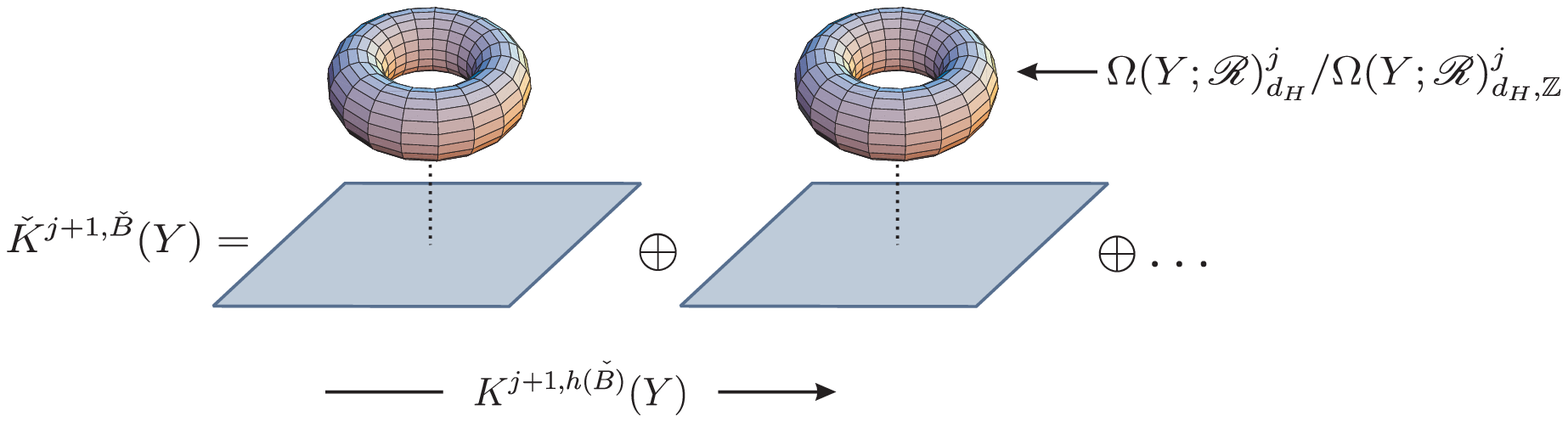}
\end{figure}

\noindent The group $\check{K}^{j+1,\check{B}}(Y)$ consists of several
connected components labeled by the characteristic class $x\in
K^{j+1,h(\check{B})}(Y)$. Each component is a torus fibration over a
vector space. The fibres are finite dimensional tori
$\mathcal{W}^{j}(Y)
=\Omega(Y;\Rc)^j_{d_H}/\Omega(Y;\Rc)^j_{d_H,\Zh}$ represented by
  topologically trivial flat gauge fields.

There is a product on twisted differential $K$-characters.
If $[\check{A}_1]\in\check{K}^{j_1,\check{B}_1}(Y)$
and $[\check{A}_2]\in\check{K}^{j_2,\check{B}_2}(Y)$ then the product
$[\check{A_1}] \cdot [\check{A_2}]$ is in $\check{K}^{j_1+j_2,\check{B}_1+\check{B}_2}(Y)$.
The characteristic class and the curvature of the product
$[\check{A_1}] \cdot [\check{A_2}]$ are
\begin{equation*}
x([\check{A_1}] \cdot [\check{A_2}])=x([\check{A_1}])\cup x([\check{A_2}])
\quad\text{and}\quad
F([\check{A_1}] \cdot [\check{A_2}])=
F([\check{A_1}])\wedge F([\check{A_2}])
\end{equation*}
respectively.

If $Y$ is odd dimensional, compact, and $\check{K}$-oriented manifold
then there is a perfect pairing $\check{K}^{j+1,\check{B}}(Y)
\times \check{K}^{j+1,\check{B}}\to U(1)$ \cite{Freed:2006ya}.
We will write this perfect pairing as $\int^{\check{K}}_Yu\check{A}_1\cdot
\phi_{j+1}(\check{A}_2)$ where $\phi_{j+1}:\check{K}^{j+1,\check{B}}(Y)
\to \check{K}^{10-j,-\check{B}}(Y)$.

In terms of differential cohomology classes the action of the
previous section is generalized to be
\begin{equation*}
e^{-S(\check{C},\check{A})}=\exp\left[ -\frac{\pi}{2}\int_X
(F(\check{C})-\check{A})\wedge *_E(F(\check{C})-\check{A}) -i\pi
\int_X^{\check{K}} u\check{C}\cdot \phi_{j+1}(\check{A}) \right]. \label{lagr2}
\end{equation*}

\paragraph{Category of twisted differential K-cocycles.}
As in Yang-Mills theory, locality forces one to work with gauge
potentials, rather than gauge isomorphism classes of fields. In
generalized abelian gauge theories the proper framework   is to find
a groupoid whose set of isomorphism classes is the set of gauge
equivalence classes. The objects in the category are the gauge
potentials and the ``gauge transformations'' are the morphisms
between objects.

In this paper we will
postulate that there exists a category
$\check{\mathscr{H}}^{j+1,\check{B}}(Y)$ which is  a
groupoid obtained by the action of a gauge group on a set of
objects.
The gauge group, from which we get the  morphisms of the category
$\check{\mathscr{H}}^{j+1,\check{B}}(Y)$ is, by hypothesis,  the group
$\check{K}^{j,\check{B}}(Y)$. There are a few ways to motivate this
definition. First, it is the natural generalization of the case of
differential cohomology, where the gauge field is a cocycle in
$\check Z^{j+1}$ and considerations of open Wilson lines
imply that the group of  gauge transformations should  $\check H^{j}$
\cite{Diaconescu:2003bm}. More fundamentally, we will adopt
Dan Freed's viewpoint \cite{Freed:2000ta} that one should begin with
a RR current $[\check j] \in \check{K}^{j+1,\check{B}}(Y)$
(induced by the background and $D$-branes, if present) and
view the space of RR fields as the set of trivializations of
$[\check j]$. The set of such trivializations is a torsor for
$\check{K}^{j,\check{B}}(X)$.
In the Chern-Simons approach, the gauge potential restricted to $X$
is identified with   the RR current. It follows that
we should identify the gauge group of the Chern-Simons theory with
$\check{K}^{j,\check{B}}(X)$. Let us now make the consequences of this
viewpoint a little more concrete.

First, the set of objects of our category  forms a space,
$\mathscr{C}(Y)$. Connected components   are labeled by
$K^{j+1,h(\check{B})}(Y)$. \textit{We assume that each component can be taken
to be a  torsor for $\Omega(Y;\Rc)^j$.}  At the cost of
naturality, we may choose a basepoint $\check{A}_\bullet$, and write
$\check{A} = \check{A}_\bullet + a$, with $a\in
\Omega(Y;\Rc)^j$. The gauge
transformations are given by
\begin{equation}
g_{\check{C}}\check{A}=\check{A}+F(\check{C}).
\end{equation}
Next, note  that flat characters $K^{j-1,h(\check{B})}(Y;\Rh/\Zh)$ act trivially
on the space of gauge fields  $\check{\mathscr{H}}^{j+1,\check{B}}(Y)$,
therefore the group of automorphisms of any object is
$\mathrm{Aut}(\check{A})=K^{j-1,h(\check{B})}(Y;\Rh/\Zh)$.

\paragraph{Twisted bundles and the  twisted Chern character.}
In this section we review some background material on twisted Chern
charactes. We follow \cite{AtiyahSegal0}. See also
\cite{Bouwknegt:2001vu,Mathai:2002yk,Freed:2002gr} for alternative
formulations.

 Elements of the
twisted differential $K$ group are represented by twisted
 bundles with connection.
Each \textit{twisted bundle} $P\to X$   arises from locally-defined
bundles of   Hilbert spaces on open covers in $X$
\cite{AtiyahSegal0,Kapustin:1999di}. So we choose an open covering
$\{X_{\alpha}\}$ of $X$ and isomorphisms
$P|_{X_{\alpha}}\cong\mathbb{P}(E_{\alpha})$ where $E_{\alpha}$ is a
Hilbert space bundle on $X_{\alpha}$. For a sufficiently fine
covering the gluing functions between the charts can be realized by
isomorphisms
\begin{equation}
g_{\alpha\beta}:E_{\alpha}|_{X_{\alpha\beta}}\to E_{\beta}|_{X_{\alpha\beta}}
\end{equation}
where $X_{\alpha\beta}=X_{\alpha}\cap X_{\beta}$. Over threefold
intersection $X_{\alpha\beta\gamma}$ the composition
$g_{\alpha\beta}g_{\beta\gamma}g_{\gamma\alpha}$ is a multiplication
by a $U(1)$ valued function
$f_{\alpha\beta\gamma}:X_{\alpha\beta\gamma}\to U(1)$. These
functions $\{f_{\alpha\beta\gamma}\}$ satisfy the cocycle condition
over fourfold intersections
$f_{\alpha\beta\gamma}f_{\beta\gamma\delta}^{-1}
f_{\gamma\delta\alpha}f_{\delta\alpha\beta}^{-1}=1$. They also
define an integral \v{C}ech cocycle  by
\begin{equation}
h_{\alpha\beta\gamma\delta}=\frac{1}{2\pi i}\bigl[
\log f_{\alpha\beta\gamma}-\log f_{\beta\gamma\delta}
+\log f_{\gamma\delta\alpha}
-\log f_{\delta\alpha\beta}
\bigr]\in\Zh.
\end{equation}
The corresponding integral cohomology class $h_P\in H^3(X;\Zh)$ is called Dixmier-Douady class.

The same structure can be rewritten in slightly different form
\cite{murray,Mathai:2003jx}:
since the twisted bundle $P$ is locally described by $\mathbb{P}(E_{\alpha})$
the Hilbert bundles $E_{\alpha}$ and $E_{\beta}$ can
differ by a line bundle $L_{\alpha\beta}$ over the twofold intersection $X_{\alpha\beta}$.
Over threefold intersections we
have to specify an isomorphism
\begin{equation}
L_{\alpha\beta}\otimes L_{\beta\gamma}\cong
L_{\alpha\gamma}\quad\text{on }X_{\alpha\beta\gamma}.
\label{3L}
\end{equation}
This isomorphism is given by multiplication by a $U(1)$ valued
function $f_{\alpha\beta\gamma}:X_{\alpha\beta\gamma} \to U(1)$. The
set of line bundles over twofold overlaps satisfying condition
\eqref{3L} is called \textit{a bundle gerb}.

\textit{A connection} on a twisted bundle $P$ locally arises from a
set of connections on the Hilbert bundles $\{E_{\alpha}\to
X_{\alpha}\}$. Let $\nabla_{\alpha}$ be a connection on the Hilbert
bundle $E_{\alpha}$. On twofold intersections these connections are
glued by
\begin{equation}
\nabla_{\alpha}|_{X_{\alpha\beta}}
=A_{\alpha\beta}+\nabla_{\beta}|_{X_{\alpha\beta}}
\label{con.glue}
\end{equation}
where $A_{\alpha\beta}$ is a connection on the line bundle $L_{\alpha\beta}\to X_{\alpha\beta}$.
Because of the condition \eqref{3L} the connections $\{A_{\alpha\beta}\}$
satisfy nontrivial relations on threefold overlaps
\begin{equation}
A_{\alpha\beta}+A_{\beta\gamma}+A_{\gamma\alpha}=
f_{\alpha\beta\gamma}^{-1}df_{\alpha\beta\gamma}\quad\text{on }X_{\alpha\beta\gamma}.
\label{3A}
\end{equation}
The set of one forms $\{A_{\alpha\beta}\}$ on twofold overlaps satisfying condition \eqref{3A}
over threefold overlaps is also known as \textit{a connecting structure}
\cite{hitchin}.

It happens that to define a differential characteristic class such
as a twisted Chern character, in addition to choosing a connection
on a twisted bundle, one must also choose a gerb connection. For a
given connecting structure \textit{a gerb connection} is described
as a set of $2$-forms $\{B_{\alpha}\}$ on charts $\{X_{\alpha}\}$
satisfying the gluing condition on twofold overlaps
\begin{equation}
B_{\beta}-B_{\alpha}=\frac{1}{2\pi i}\,dA_{\alpha\beta}\quad\text{on }X_{\alpha\beta}.
\label{B.glue}
\end{equation}
From this condition it follows that there exists a globally well
defined closed $3$-form $H\in\Omega^3_{\Zh}(X)$ (locally
$H|_{X_{\alpha}}=dB_{\alpha}$) whose de Rham cohomology class
$[H]_{dR}$ coincides with the image of integral cohomology class
$h_P$ under the natural projection to  de Rham cohomology.

\textit{A twisted Chern character} $\ch_{\check{B}}(P)$ arises locally
from a connection on Hilbert bundles over $\{X_{\alpha}\}$ and a gerb connection $\check{B}$:
\begin{equation}
\ch_{\check{B}}(P)|_{X_{\alpha}}=e^{B_{\alpha}}\ch(\nabla_{\alpha})
\label{chB}
\end{equation}
where $\ch(\nabla_{\alpha})$ is the Chern character for the Hilbert
bundle $E_{\alpha}$. (If we write an expression involving a trace
then that trace needs to be regulated.)  If $E_{\alpha}$ is finite
dimensional \footnote{In general, we have to consider a
$\Zh_2$-graded  twisted bundle. The Chern character is defined by
$\mathrm{str}\,e^{\mathcal{F}/2\pi i}$ where $\mathcal{F}$ is the
curvature of a superconnection and $\mathrm{str}$ is the super
trace.} then $\ch(\nabla_{\alpha})=\tr e^{F(\nabla_{\alpha})/2\pi
i}$. The gluing conditions \eqref{con.glue} and \eqref{B.glue}
ensure that \eqref{chB} is globally well defined.

\paragraph{$B$-field gauge transformations.}
In the previous paragraph we described a twisted bundle with
connection locally using Hilbert bundles with connection. This local
structure has a nontrivial group of automorphisms given by
$\check{H}^2(X)$ (the group of isomorphism classes of line bundles
with connection). Indeed given a globally well defined line bundle
$L\to X$ with connection $\nabla$ we can change each Hilbert bundle
$E_{\alpha}$ to $E_{\alpha}\otimes L|_{X_{\alpha}}$ and
$\nabla_{\alpha}\mapsto \nabla_{\alpha}+\nabla|_{X_{\alpha}}$.
Notice that all equations but \eqref{chB} of the previous paragraph
remains unmodified. So to preserve equation \eqref{chB} we also need
to change the gerb connection by $B_{\alpha}\mapsto
B_{\alpha}-\frac{1}{2\pi i} F(\nabla)|_{X_{\alpha}}$ where
$\frac{1}{2\pi i} F(\nabla)\in \Omega^2_{\Zh}(X)$.   This
automorphism is called \textit{$B$-field gauge transformation}.

One must distinguish between $B$-field gauge transformations and a
change of the $B$-field: the $B$-field gauge transformation is
\textit{a two step} transformation described above and defined in
such a way that it preserves the twisted Chern character. By
contrast, under a \textit{change} of the $B$-field: $\check{B}$ goes
to $\check{B}+b$ for $b\in\Omega^2(X)$, the twisted Chern character
changes by
\begin{equation}
\mathrm{ch}_{\check{B}+b}(P)=e^{u^{-1}b}\,\mathrm{ch}_{\check{B}}(P).
\end{equation}

\paragraph{Chern classes.} The Chern classes and the Chern character for
twisted bundles are related in an unusual fashion. Before discussing
this in   detail we consider a simple example. Suppose we have a
topologically trivial $B$-field in type IIA supergravity. It is then
possible   to solve the Bianchi identity for the Ramond-Ramond field
$R$ in the form
\begin{equation}
R=e^{u^{-1}B} F
\label{RBF}
\end{equation}
where $F$ is a \textit{closed} form of even degree,
$F=F_0+u^{-1}F_2+\dots+u^{-5}F_{10}$.  In the previous paragraph we
explained that besides the physical data (= information about the
twisted bundle $P$) $F$ contains auxiliary data which is cancelled
in \eqref{RBF} by the $B$-field. In other words $F$ describes a
Hilbert bundle $E$ rather than a twisted one. When we apply an
automorphism $F$ changes to $e^{u^{-1}\omega}F$ and $B\mapsto
B-\omega$. This means that we have to keep track of a flat
$B$-field. To separate these auxiliary data we should consider
invariant polynomials $c(F)$ such that $c(e^{u^{-1}\omega}F)=c(F)$.
Such polynomials form a ring, and Chern classes can be taken to be
generators of this ring \cite{AtiyahSegal}.

\vspace{0.5cm} \example{For example, if $F_0\ne 0$ then the
 invariant
ring is generated by the polynomials
\begin{equation*}
F_0,\quad 2F_0F_4-F_2^2,\quad 3F_0^2F_6-3F_0F_2F_4+F_2^3,
\quad 2F_0F_8-2F_2F_6+F_4^2,
\quad 4F_0^2F_4^2-4F_4F_2^2F_0+F_2^4,\quad\text{etc}.
\end{equation*}
So there is one generator in degrees $1$, $4$ and $6$ and two
generators in degree $8$ and $10$. The Poincar\'{e} polynomial for
this ring was calculated in \cite{AtiyahSegal}. If $F_0=0$ then the
generators are given by the polynomials above after the substitution
$F_0\to F_2$, $F_2\to F_4$, etc. Notice that in this case there is a
significant reduction in the number of invariants.

For type IIB the ring is generated by
\begin{equation*}
F_1,\quad F_1F_3,\quad F_1F_7-F_3F_5,\quad F_1F_3F_5,\quad\text{etc}.
\end{equation*}

The RR topological sectors of supergravity
must be specified in terms of invariant polynomials.
}

Locally Chern classes come from invariant polynomials in the
curvature $F_{\alpha}:=F(\nabla_{\alpha})$ of a connection on a
Hilbert bundle $E_{\alpha}\to X_{\alpha}$. Consider the monomials
$x_0=\tr \mathbf{1},\, x_2=\tr F_{\alpha},\,x_4=\frac12\tr
F_\alpha^2,\dots$ and $x=x_0+u^{-1}x_2+u^{-2}x_4+\dots$ (for a
suitably defined trace). Denote by $\mathcal{J}_{\alpha}\to
X_{\alpha}$ the ring of polynomials in $x_0,\,x_2,\,x_4,\dots$ which
are invariant under $x\mapsto e^{u^{-1}\omega}x$ where $\omega$ is a
closed $2$-form. It is easy to see that the rings of invariant
polynomials $\{\mathcal{J}_{\alpha}\}$ glue together into a ring
$\mathcal{J}$ of invariants over $X$. The generators of this ring
are the rational Chern classes.

\clearpage
\section{Defining Chern-Simons terms in 11 dimensions}
\label{sec:defCSAB}\setcounter{equation}{0}

\paragraph{Chern-Simons functional}
The Chern-Simons functional for  the differential $K$-group was
defined by Hopkins and Singer \cite{Hopkins:2002rd}. In this paper
we are using a generalization of their functional to the case of a
\textit{twisted} differential $K$-group. Gauge inequivalent fields
are elements of twisted differential $K$-group:  $[\check{A}]\in
\check{K}^{j+1,\check{B}}(Y)$.  Notice that $[u\check{A}\cdot
\phi_{j+1}(\check{A})]$ is an element of the untwisted group
$\check{K}^0(X)$ and moreover it has a natural lift to the
differential $KO$-group
\cite{Hopkins:2002rd,Freed:2000ta,Freed:prv}. The Chern-Simons
functional $\CSK$ is defined by
\begin{equation}
e^{2\pi i \CSK^{j+1,\check{B}}_Y(\check{A})}:=
\exp\Bigl[i\pi \int^{\check{K}O}_Yu\check{A}\cdot \phi_{j+1}(\check{A})
\Bigr].
\label{CSK-HS}
\end{equation}
Here $\int^{\check{K}O}_X:
\check{K}O^0(X)\to \check{K}O^{-\dim X}(pt)$ denotes integration in the differential
$\check{K}O$-theory where
\begin{center}
\renewcommand{\arraystretch}{1.3}
\begin{tabular}{|c||c|c|c|c|c|c|c|c|}
\hline
\rowcolor[gray]{0.85}
$j\mod 8$& $0$ & $1$ & $2$ & $3$ & $4$ & $5$ & $6$ & $7$
\\
\hline $\check{K}O^{-j}(\mathrm{pt})$ & $\;\,\Zh\;\,$ & $\Zh/2$ &
$\Zh/2$ & $\Rh/\Zh$ & $\;\,\Zh\;\,$ & $\;\,0\;\,$ & $\;\,0\;\,$ &
$\Rh/\Zh$
\\
\hline
\end{tabular}
\end{center}
The groups which are important for our case correspond to $j$ equal
to $3$ and $2$. $\check{K}O^{-3}(pt)\cong \Rh/\Zh$ defines for us
the Chern-Simons functional as in \eqref{CSK-HS}. The integral is
essentially given by the $\eta$-invariant $+$ local Chern-Simons
term \cite{Freed:2006ya}. The group $\check{K}O^{-2}(pt)\cong \Zh/2$
will play a crucial role in formulating the Gauss law. The integral
in this case is given by the mod two index of the Dirac operator.
Thus the $\Omega$-function of Witten \cite{Witten:1999vg} is
automatically included in the definition of the $K$-theoretic
Chern-Simons functional.

From \eqref{CSK-HS} it follows that the Chern-Simons functional
is \textit{a quadratic refinement of the bilinear form}
on $\check{K}^{j+1,\check{B}}\times \check{K}^{j+1,\check{B}}$:
\begin{equation}\label{fancycs}
\CSK(\check{A}_1+\check{A}_2)=
\CSK(\check{A}_1)+\CSK(\check{A}_2)
+\int^{\check{K}}_Y
u\check{A}_1\cdot\phi_{j+1}(\check{A}_2)
\mod 1
\end{equation}

\paragraph{Coupling to RR current and $D$-branes.}

The coupling to RR current and $D$-branes is easily
included in the above formalism.  The
data specifying a wrapped D-brane state includes a choice (of a
pair) of a twisted bundle $P\to W$ with connection. Locally a twisted
bundle is described by a projective Hilbert bundle (for more details
see section~\ref{sec:fieldspace} and \cite{AtiyahSegal}). The
charges of the brane are described in terms of  the topology of the
embedded cycle $f: W\hookrightarrow X$ and the topology of $P$.
In addition, the twisting of the normal bundle $\nu\to W$ contributes to the
RR current    \cite{Cheung:1997az, Minasian:1997mm, Freed:1999vc}.
We denote by $\check{\tau}$ a differential
refinement of the third integral Wu class $W_3(\nu)$ on $\nu$. There
is a natural integral cohomology class $h_P\in H^3(W;\Zh)$
associated to the twisted bundle $P$
--- Dixmier-Douady class.
The Freed-Witten anomaly cancellation condition \cite{Freed:1999vc}
requires that $h_P=h(\check{\tau}+f^*\check{B})=0$ where
$f^*\check{B}$ is the pullback of the $B$-field $\check{B}$ from
$X$. \footnote{This is the anomaly cancellation condition which
should properly be referred to as the ``Freed-Witten anomaly.'' The
famous shift of the $U(1)$ flux quantization by $\frac12 c_1$ for
$D$-branes wrapping non-spin manifolds was already pointed out
earlier in \cite{Minasian:1997mm}. A very useful interpretation of
all these facts (and more) was advocated
 by D. Freed in \cite{Freed:2000ta} where the $U(1)$ gauge degree of
freedom was interpreted as a trivializing cochain in differential
cohomology. }

The twisted bundle $P$ defines a class $[P]$ in the twisted
$K$-group $K^{0,h_P}(W)$. The  cohomology class of the RR current
created by the   D-brane is a pushforward of $[P]$ to the twisted
$K$-group on $X$
\cite{Minasian:1997mm,Moore:1999gb,Bouwknegt:2000qt}
\begin{equation}
\mathrm{ch}_{\check{B}}(f_![P])\sqrt{\hat{A}(X)}
\end{equation}
where $f_!$ is the K-theoretic Gysin map,
$f_!:K^{0,[f^*\check{B}+\check{\tau}]}(W)\to K^{ -\mathrm{codim}\,
W,[\check{B}]}(X)$. There is a generalization of this equation to
the differential $K$ group \cite{Hopkins:2002rd} allowing one to
define the RR \textit{current} associated to the $D$-brane. In
general the RR current   is   as an element $[\check{j}]$ of the
twisted differential $K$-group $\check{K}^{j+1,\check{B}}(X)$. If
the current is extended to  $[\check{j}]\in K^{j+1,\check{B}}(Y)$ we
can consider the coupling to $[\check{A}]$ to be of the form
\begin{equation}
\exp\left[2\pi i \int_Y u\check{A}\cdot\phi_{j+1}(\check{j})\right].
\label{Ccopuling2}
\end{equation}
Note that from \eqref{fancycs}, we see that by shifting
$\check{A}\mapsto \check{A}-\check{j}$ we can cancel the source term
in \eqref{Ccopuling2}.

\paragraph{Variational formula.}
Suppose we are given a family $Z$ of $\check{K}O$-oriented
$11$-manifolds over
the
\begin{wrapfigure}{l}{110pt}
\includegraphics[width=100pt]{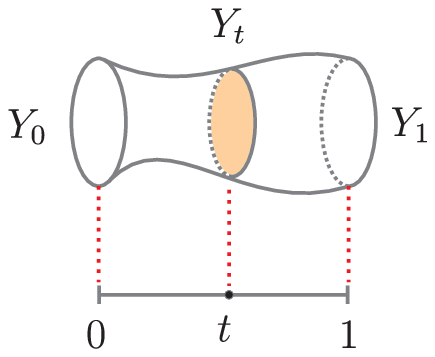}
\end{wrapfigure}
interval $[0,1]$. We denote by $Y_t$ the fibre of this family over
the point $t$. Let $\check{A}\in \check{K}^{j+1,\check{B}}(Z)$ be a
differential cocycle. Then one can show that
\begin{equation}
\CSK_{Y_1}^{j,\check{B}}(\check{A}_1)
-\CSK_{Y_0}^{j,\check{B}}(\check{A}_0) =\frac{1}{2}\int_{Z}
u F(\check{A})\wedge \phi_{j+1}(F(\check{A}))
\mod 1
\label{variat}
\end{equation}
where $\check{A}_0$ and $\check{A}_1$ denote restrictions
of $\check{A}$ to $Y_0$ and $Y_1$ respectively.

\paragraph{Chern-Simons functional on lower dimensional manifolds.}
On a compact $11$-dimensional $\check{K}O$ oriented manifold $Y$ the
Chern-Simons functional defines a map from the space of gauge fields
on $Y$ to $\Rh/\Zh$:
\begin{subequations}
\begin{equation}
\CSK_Y:\{\text{gauge fields on Y}\}\to \Rh/\Zh.
\end{equation}
Actually the same Chern-Simons functional makes sense on lower
dimensional manifolds. On a compact $10$-dimensional $\check{K}O$
oriented manifold $X$ Chern-Simons functional can be used to define
a line bundle with connection over the space of gauge fields on $X$
--- an element of $\check{H}^2\bigl(\frac{\{\text{gauge fields on
$X$}\}}{\{\text{gauge tr.}\}} \bigr)$.
%
%
\end{subequations}

\section{Quantization of ``spin'' Chern-Simons theory}
\label{sec:CSACSB}\setcounter{equation}{0}
In general,
there   are two ways to quantize Chern-Simons theory: one can first
impose the equation of motion  classically and then quantize the
space  of solutions of this equation, alternatively one can first
quantize the space of all gauge fields and then impose the equation
of motion as an operator constraint. In this paper we mostly follow
the second approach, although our ultimate goal is to construct
wavefunctions on the gauge invariant phase space.

Consider the following topological field theory on an $11$-dimensional
manifold $Y$
\begin{equation}
e^{2\pi i \CSK_Y^{j+1,\check{B}}(\check{A})}=
\exp\Bigl[
i\pi \int_Y^{\check{K}O}u\check{A}\cdot\phi_{j+1}(\check{A})
\Bigr].
\label{top_action}
\end{equation}
Note that the Chern-Simons functional is \textit{not} necessarily symmetric:
there exist $\check{\lambda}\in \check{K}^{j+1,\check{B}}(Y)$
such that $\CSK(\check{\lambda}-\check{A})=\CSK(\check{A}) \mod 1$.
This equation actually fixes only the characteristic class $\lambda$ of $[\check{\lambda}]$.
It happens that on $10$-manifold $\lambda_X$ is always divisible by $2$ in the
twisted $K$-theory. The Chern-Simons functional defines a preferred
class $\mu\in K^{j+1,h(\check{B})}(X)$ such that $\lambda_X=2\mu$
\cite{Hopkins:2002rd}.

 Using the variational formula
\eqref{variat} one obtains the familiar equation of motion
\begin{equation}
F(\check{A})=0.
\label{eqr}
\end{equation}
One might think of $Y$ as a product space $\Rh\times X$ and proceed
with Hamiltonian quantization of this theory. However we proceed
differently: in the previous section we said that Chern-Simons on a
$10$-manifold $X$ defines a line bundle with connection over the
space of gauge fields on $X$. In this section we describe this
construction in detail.



\paragraph{CS line bundle and connection.}
The space of the gauge fields is an infinite-dimensional
 space of objects of the category $\check{\mathscr{H}}^{j+1,\check{B}}(X)$:
\begin{equation*}
\mathscr{C}(X):=\mathrm{Obj}(\check{\mathscr{H}}^{j+1,\check{B}}(X)).
\end{equation*}
The topological action \eqref{top_action} defines a natural line
bundle $\mathcal{L}\to \mathscr{C}(X)$ (the Chern-Simons line
bundle). Chern-Simons functional on an $11$-manifold with boundary
$X$ is most naturally considered as a section of $\mathcal{L}$.

The Chern-Simons line bundle $\mathcal{L}$ has a natural connection defined
as follows: Consider a
path $\check{A}_t$ in the space of twisted differential
$K$-cocycles  $\mathscr{C}(X)$ where $t\in[0,1]$ is the coordinate on the
path. One can think of $\check{A}_t$ as of a twisted differential $K$-cocycle
from $\mathscr{C}([0,1]\times X)$.
The parallel transport is defined by
\begin{equation}
\mathscr{U}(\{\check{A}_t\}):=e^{2\pi i \CSK^{j+1,\check{B}}_{[0,1]\times X}(\check{A}_t)}
\in\mathrm{Hom}(
\Lc|_{\check{A}_0},\Lc|_{\check{A}_1}).
\label{connection}
\end{equation}
The tangent vector to the path $\{\check{A}_t\}$
is $\xi\in\Omega(X;\Rc)^j$. The curvature of the connection
\eqref{connection} can be computed from the variational formula
\eqref{variat}:
\begin{equation}
\Omega_{\check{A}_X}(
\xi_1,\xi_2)=-2\pi i \,
\omega_j(\xi_1,\xi_2)
\end{equation}
where $\omega_j$ is the symplectic form \eqref{omegaj}.

\begin{wrapfigure}{l}{95pt}
\vspace{-3mm}
\includegraphics[width=90pt]{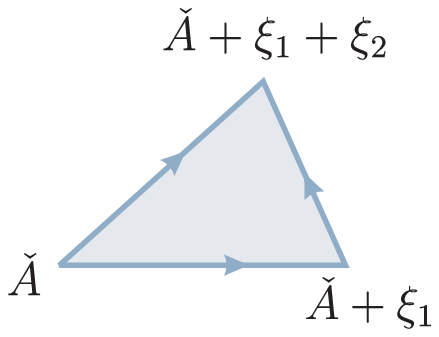}
\end{wrapfigure}
\noindent For any $\xi\in\Omega(X;\Rc)^j$ we
introduce a straightline
path $p_{\check{A};\xi}(t)=\check{A}+t\xi$
in the space of
twisted differential $K$-cocycles  $\mathscr{C}(X)$.
Using the variational formula \eqref{variat} one finds
\begin{equation}
\mathscr{U}(p_{\check{A}+\xi_1;\xi_2})
\mathscr{U}(p_{\check{A};\xi_1})
=e^{-i\pi \omega_j(\xi_1,\xi_2)}\,\mathscr{U}(p_{\check{A};\xi_1+\xi_2})
.
\end{equation}

\noindent Now we need to lift the action of the gauge group (defined in
section \ref{sec:fieldspace} above) to the line bundle $\mathcal{L}$.
\begin{wrapfigure}{l}{150pt}
\includegraphics[width=140pt]{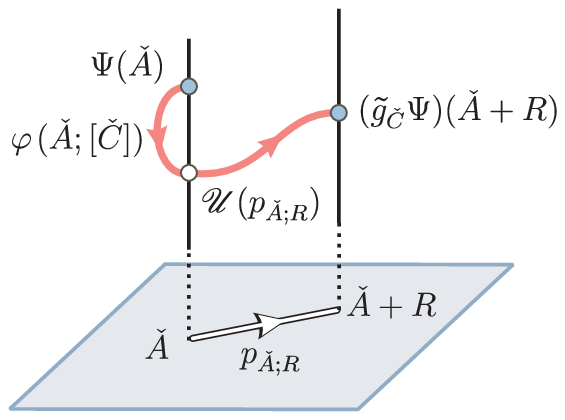}
\end{wrapfigure}
The difference between a
group lift and parallel transport is a cocycle. That is, we can
define the group lift by
\begin{equation}
(\tilde{g}_{\check{C}}\Psi)(g_{\check{C}}\check{A})
:=\varphi(\check{A};[\check{C}])
\mathscr{U}(p_{\check{A};R(\check{C})})
\Psi(\check{A})
\label{Gaugetr}
\end{equation}
provided  $\varphi$ is a  phase satisfying the cocycle condition:
\begin{multline}
\varphi(g_{\check{C}_1}\check{A};[\check{C}_2])
\varphi(\check{A};[\check{C}_1])
\\
=\varphi(\check{A};[\check{C}_1]+[\check{C}_2]) \,e^{-i\pi\,
\omega_j(R(\check{C}_1),R(\check{C}_2))}. \label{cocycle}
\end{multline}

Once the group lift is defined we can impose an operator constraint
on the wave function --- the   Gauss law --- which says
$\tilde{g}_{\check{C}}\cdot \Psi(\check{A})=
\Psi(g_{\check{C}}\check{A})$ (for more details see
section~\ref{subseq:QGL}).

\subsection{Construction of the cocycle via a Chern-Simons term }
One way to construct a cocycle proceeds using a construction going
back to Witten and described in detail in  \cite{Diaconescu:2003bm}.

 Chern-Simons functional provides a natural candidate
 for the  cocycle $\varphi$ \cite{Witten:1999vg,Diaconescu:2003bm}.
Construct a differential cocycle on the closed $11$-manifold
$S^1\times X$:
\begin{equation*}
\check{A}_X+\check{t}\cdot\check{C}
\end{equation*}
where $[\check{t}]\in \check{K}^{-1}(S^1)$ is a differential
$K$-character with field strength $F(\check{A})+dt\wedge R(\check{C})$
and characteristic class $x(\check{A})+[dt]\cup x(\check{C})$.
\footnote{If we choose a noncanonical splitting
$\check K^1(S^1) \cong \Rh/\Zh \times \Omega^1_{\Zh}(S^1)$ then
we can take $[\check t] = (dt, 0)$. }

Now using this twisted differential
character we can define
\begin{equation}
\varphi(\check{A}_X;[\check{C}]):=
e^{2\pi i \CSK^{j+1,\check{B}}_{S^1\times X}(\check{A}_X+\check{t}\cdot \check{C})}.
\label{cocycle_def}
\end{equation}
To define a Chern-Simons term we need to choose a
$\check{K}O$-orientation for $S^1\times X$: it includes in
particular a choice of spin structure on the $S^1$, which should be
the bounding spin structure $S^1_-$. A standard cobordism argument
(see, for examples \cite{Diaconescu:2003bm,Belov:2006jd} ) shows
that the functional \eqref{cocycle_def} satisfies the cocycle
relation \eqref{cocycle}.

Using properties of the multiplication of differential characters one
can rewrite the cocycle \eqref{cocycle_def} as
\begin{equation}
\varphi(\check{A}_X;[\check{C}])=
\Omega([\check{C}])\,e^{2\pi i \int_X [u\check{C}]\cdot [\phi_{j+1}(\check{A}_X)]}
\label{newomega}
.
\end{equation}
{}It follows that
$\Omega([\check{C}])$ is a locally constant function of
$[\check{C}]$. Therefore it only depends on the characteristic class $x([\check{C}])$. Since
there is no difficulty in defining the integral level Chern-Simons
term,   $\Omega$ must take values $\{\pm
1\}$. Finally, from the cocycle condition we derive:
\begin{equation}
\Omega([\check{C}_1]+[\check{C}_2])=\Omega([\check{C}_1])\,
\Omega([\check{C}_2])(-1)^{\omega_j(x([\check{C}_1])
,x([\check{C}_2]))}.
\label{eqOmega1}
\end{equation}

\paragraph{QRIF.} A function $\Omega:K^{j+1,h(\check{B})}(X)\to U(1)$
which satisfies the quadratic equation \eqref{eqOmega1} is called a
QRIF. QRIF stands for  Quadratic Refinement of the Intersection
Form.

Now, associated to $\Omega$ is an important invariant. Note that
since the  bilinear form $\omega_j(x,y)$ vanishes on   torsion
classes, $\Omega$ is a homomorphism from $\mathrm{Tor}\,K^{j+1,h(\check{B})}(X)$
to $\Rh/\Zh$. Since there is a perfect pairing on torsion classes it
follows that there is a $\mu \in\mathrm{Tor}\,K^{j+1,h(\check{B})}(X)$ such
that
\begin{equation}
\Omega(x_T) = e^{2\pi i\, T( x_T, \mu ) } = e^{2\pi i \,
\int_X u\alpha \cup
\phi_{j+1}(\mu) } \label{mu}
\end{equation}
for all torsion classes $x_T$.

As in our previous paper, if one starts with an $\Omega$ function,
as in \cite{Witten:1999vg}, then there appear to be many choices.
Within the Chern-Simons framework there is a distinguished choice
for $\Omega$ following from \eqref{cocycle_def}. This choice of
$\Omega$, which is essentially the mod two index, has been claimed
to be the unique choice compatible with $T$-duality
\cite{Witten:1999vg,Moore:2002cp}. Moreover, it appears to be the
unique choice that is compatible with $M$-theory
\cite{Diaconescu:2000wy,Diaconescu:2000wz}, so we will make that
choice from now on. \footnote{In \cite{Freed:2004yc} the possibility
was left open that one  could possibly add $\Zh_2$-valued
topological terms to the $M$-theory action. If that is indeed
possible then there would be a corresponding modification of
$\Omega$. If that possibility can be ruled out, then the choice of
$\Omega$ based on the mod-two index is unique. }

\subsection{Quantum Gauss law}
\label{subseq:QGL} The wave function must define a section of the
line bundle $\Lb$ over a component of the space of
gauge equivalence classes of fields $\check{K}^{j+1,\check{B}}(X)$.
In the previous section we constructed a line bundle
$\mathcal{L}$ over the objects $\mathscr{C}(X)$ of the category
$\check{\mathscr{H}}^{j+1}(X)$. A
section $\Psi$ of $\mathcal{L}\to \mathscr{C}$ descends to a section
of $\Lb$ iff it satisfies the Gauss law constraint
\begin{equation}
\Psi(g_{\check{C}}\check{A})=\varphi(\check{A};[\check{C}])\,
\mathscr{U}(p_{\check{A};R(\check{C})})\Psi(\check{A}).
\label{Glaw}
\end{equation}

\paragraph{Tadpole constraint.}
Equation \eqref{Glaw} has solution only on \textit{one} connected
component of $\check{K}^{j+1,\check{B}}(X)$. Recall that the objects
in $\mathscr{C}(X)$ have a nontrivial group of automorphisms:
\begin{equation*}
\mathrm{Aut}([\check{A}])\cong K^{j-1,h(\check{B})}(X;\Rh/\Zh).
\end{equation*}
It
is easy to see that \eqref{Glaw} admits a nonzero solution only if
the automorphism group of an object acts trivially. This amounts to
the condition
\begin{equation*}
\varphi(\check{A};[\check{C}])=1
\quad\text{for}\quad
[\check{C}]\in K^{j-1,h(\check{B})}(X;\Rh/\Zh).
\end{equation*}
Combining \eqref{newomega}
with \eqref{mu} we obtain the ``tadpole condition'':
\begin{equation}
\mu + x(\check{A})=0.
 \label{tadpole}
\end{equation}

\paragraph{Cocycle.}
The cocycle $\varphi$ looks particularly simple for differential
characters satisfying the tadpole constraint \eqref{tadpole}:
Suppose we are given a $\Zh_2$-valued QRIF $\Omega_0(x)$ with
vanishing torsion class $\mu$ (we will specify such a QRIF, defined
by a certain Lagrangian decomposition, in the next section).
Actually it depends only on the rational $K$-theory class, so we
will also write it as $\Omega_0([R]_{d_H})$. Then there exists a
twisted differential $K$-character $[\check{\mu}]$ such that
\begin{equation}
\Omega([\check{C}])=\Omega_0([R]_{d_H})\,e^{2\pi i \int_X u[\check{C}]
\cdot \phi_{j+1}([\check{\mu}])}
\end{equation}
where $R$ is the curvature of the character $[\check{C}]$. Moreover,
given a choice of $\Omega_0$, $[\check \mu]$ is uniquely determined
by this equation since the quantity on the right-hand side is a
perfect pairing. Note that  the characteristic class
$x([\check{\mu}])$ is exactly $\mu$. Since $\Omega([\check{C}])$ is
$\Zh_2$-valued $[\check{\mu}]$ must be $2$-torsion. In particular,
the fieldstrength of $\check{\mu}$ is zero.

 Now we can rewrite the phase in \eqref{Glaw}
in the form
\begin{equation}
\varphi(\check{A};[\check{C}])=
\Omega_0([R]_{d_H})
\,e^{2\pi i
\int_X u[\check{C}]\cdot \phi_{j+1}([\check{A}]+[\check{\mu}])}.
\label{varphi-final}
\end{equation}
For
$\check{A}$ satisfying the tadpole constraint \eqref{tadpole} the
character $[\check{\sigma}]=[\check{A}]+[\check{\mu}]$ is
topologically trivial. Thus it can be
trivialized by a form $\sigma(\check{A}+\check{\mu})$
of total degree $j$ which satisfies two properties
\begin{equation}
d_H\sigma(\check{A}+\check{\mu})=F(\check{A})
\quad\text{and}\quad
\sigma(\check{A}+\check{\mu}+a)=\sigma(\check{A}+\check{\mu})+a
\quad\forall a\in\Omega(X;\Rc)^{j}.
\label{sigmadef}
\end{equation}

So finally the Gauss law \eqref{Glaw} can be written as
\begin{equation}
\Psi(g_{\check{C}}\check{A})=\Omega_0([R]_{d_H})\,e^{-2\pi i
\,\omega_j(R,\sigma(\check{A}+\check{\mu}))}
\mathscr{U}(p_{\check{A};R(\check C)})\Psi(\check{A}). \label{GlawF}
\end{equation}

\paragraph{Gauss law in local coordinates.}
Each component in the space of objects in
$\check{\mathscr{H}}^{j+1,\check{B}}(X)$ is a contractible space.
Thus the line bundle $\Lc\to \mathscr{C}$ is trivializable. To
construct a section explicitly we need to choose an explicit
trivialization of this line bundle. To this end we choose an
arbitrary twisted differential $K$-cocycle $\check{A}_{\bullet}$
satisfying the tadpole constraint \eqref{tadpole} \footnote{If
$\mu=0$ then there exists a preferred base point
$\check{A}_{\bullet}=0$. However even in this case we would like to
keep $\check{A}_{\bullet}$ arbitrary because we will use it later to
specify the external current.}. Then an arbitrary field
configuration (in the connected component) can be parameterized by
$\check{A}=\check{A}_{\bullet}+a$ where $a$ is a globally well
defined form of total degree $j$. Define a canonical nowhere
vanishing section $S$ of unit norm by
\begin{equation*}
S(\check{A}):=
\mathscr{U}(p_{\check{A}_{\bullet};\check{A}-\check{A}_{\bullet}})\, S_{\bullet}
\end{equation*}
where $S_{\bullet}\in \Ch$ and $|S_{\bullet}|=1$. The \textit{wave function}
$\mathcal{Z}(a)$
is a ratio of two sections $\Psi(\check{A})/S(\check{A})$.

{}From equations \eqref{Gaugetr} and \eqref{GlawF} it follows
that the Gauss law  for the wave function is
\begin{equation}
\mathcal{Z}(a+R)=\Omega_0([R])\,
e^{-2\pi i\,\omega_j(R,\sigma(\check{A}_{\bullet}+\check{\mu}))
-i\pi \,\omega_j(R,a)}\,
\mathcal{Z}(a)
\label{GSL2}
\end{equation}
for an arbitrary $d_H$-closed form
$R$ of total degree $j$ with integral periods.

It is clear that both the partition function $\mathcal{Z}(a)$ and
Gauss law \eqref{GSL2} depend on a choice of the base point.
So the discussion in section~5.3 in \cite{Belov:2006jd}
should be repeated.

\section{Construction of the partition function}
\label{sec:partfunc}\setcounter{equation}{0} The content of this
section is as follows: To obtain a quantum Hilbert space we need to
choose a polarization on the phase space
$P=\mathscr{C}(X)/\mathscr{G}$. A choice of Riemannian metric $g_E$
on $X$ defines a complex structure $J$ on $T_{\check{A}}P$. The
quantum Hilbert space consists of holomorphic sections $\{\Psi\}$ of
$\Lc$.

Note that there are infinitely many sections of $\mathcal{L}$
which satisfy the Gauss law \eqref{Glaw}, in contrast there are
finitely many \textit{holomorphic} sections which satisfy the Gauss
law \eqref{Glaw}. By choosing a local coordinate system
$(\check{A}_{\bullet},S(\check{A}))$ on $\Lc\to \mathscr{C}$ one can
try to construct a holomorphic solution of the Gauss law explicitly.
The resulting expression will in addition depend on some extra
choices such as a Lagrangian decomposition
$K^{j,h(\check{B})}(X)/\text{torsion}=\bar{\Gamma}_1\oplus \bar{\Gamma}_2$ of the
twisted K group  modulo torsion. The (local) expression for the partition
function is summarized by Theorem~\ref{thm:1}.

\subsection{Choice of polarization}
Equation \eqref{tadpole} constrains the  connected component in the
space of the gauge fields $\mathscr{C}$. Now by choosing a local coordinate system
$(\check{A}_{\bullet},S)$ we can identify the phase space with the real
vector space $V_{\Rh}=\Omega(X,\Rc)^j$ by $\check{A}=\check{A}_{\bullet}
+a$, $a\in V_{\Rh}$.

The vector space $V_{\Rh}$ has a natural antisymmetric form
defined by \eqref{omegaj}.
This $2$-form is closed and nondegenerate and thus it defines
a symplectic structure on the space of gauge fields $\mathscr{C}$.

Recall that a choice of Riemannian metric $g_E$ on $X$ defines
a compatible complex structure $J$, \eqref{Jdef}.
Using this complex structure we decompose the space of real forms $V_{\Rh}$ as
\begin{equation}
V_{\Rh}\otimes \Ch\cong V^+\oplus V^-.
\end{equation}
Any vector $R^{+}$ of the complex vector space $V^+$ can be
\textit{uniquely} written as
\begin{equation}
R^+=\frac12(R-iJR)
\label{Rplus}
\end{equation}
for some real vector $R\in V_{\Rh}$.

This decomposition introduces complex coordinates on the patch
$(\check{A}_{\bullet},S)$. Recall that in \textit{real} local
coordinates we have a covariant derivative $D:=\delta  -i\pi
\,\omega_j(\delta a,a)$ which is defined on sections of the line
bundle $\Lc$. Here $\delta$ is the usual differential with respect
to $a$. In   \textit{complex} coordinates the covariant
derivative $D$ decomposes as $D=D^++D^-$ where
\begin{equation}
D^+=\delta^+-i\pi\,\omega_j(\delta a^+,a^-)
\quad\text{and}\quad
D^-=\delta^--i\pi\,\omega_j(\delta a^-,a^+).
\end{equation}
The quantum Hilbert space consists of holomorphic sections, i.e.
$D^-\Psi=0$ which satisfy the Gauss law.

In the local coordinates $(\check{A}_{\bullet},S)$ one can identify
holomorphic sections $D^-\mathcal{Z}(a^+,a^-)=0$ with holomorphic functions $\vartheta(a^+)$
via
\begin{equation}
\mathcal{Z}(a^+,a^-)=e^{i\pi \,\omega_j(a^-,a^+)}\vartheta(a^+).
\end{equation}
In this case the Gauss law constraint \eqref{GSL2} takes the following simple
form
\begin{equation}
\vartheta(a^++R^+)=\bigl\{\Omega_0([R]_{d_H})
\,e^{-2\pi i\, \omega_j(R,\sigma(\check{A}_{\bullet}+\check{\mu}))}\bigr\}\,
\,e^{\frac{\pi}{2}\,H(R^+,R^+)
+\pi H(a^+,R^+)}
\,\vartheta(a^+)
\label{theta}
\end{equation}
for all $R\in\Omega(X;\Rc)^j_{d_H,\Zh}$. Here we have introduced a
hermitian form $H$ on $V^+\times V^+$. It is defined  using the
Riemannian metric $g_j$ and symplectic form $\omega_j$:
\begin{equation}
H(v^+,w^+):=2i\,\omega_j(v^+,\overline{w^+})
=g_j(v,w)+i\omega_j(v,w).
\label{H}
\end{equation}
In our notation $H$ is $\Ch$-linear in the first argument and
$\Ch$-antilinear in the second: $H(v,w)=\overline{H(w,v)}$.

\subsection{Partition function}
\label{subsec:part}

Equation \eqref{theta} looks like a functional equation for a theta
function. The important difference is that the equation for a theta
function is usually defined on a finite dimensional vector space,
while our equation is on the infinite dimensional vector space
$\Omega(X;\Rc)^j$. We solve it in a manner parallel to the discussion in
\cite{Belov:2006jd}. To write an explicit expression we must choose
a maximal isotropic (i.e. Lagrangian) subspace $V_2\subset V_{\Rh}$. This allows us
to define a    $\Ch$-bilinear form on $V^+\times V^+$ which extends $H$ from
``half'' of the space to $V_{\Rh}$. A Lagrangian subspace
 defines an orthogonal coordinate system on $V_{\Rh}$:
$V_{\Rh}=V_2\oplus JV_2$.
So any vector $v\in V_{\Rh}$ has coordinates
$v_2\in V_2$ and $v_2^{\perp}\in JV_2$.
In terms of this notation $B$ is defined by the equation
\cite{Belov:2006jd}:
\begin{subequations}
\begin{align}
(H-B)(\xi^+,\eta^+)&=2i\, \omega_j(\xi,\,\mathcal{F}^-(\eta))
\quad\text{where}\quad \mathcal{F}^-(\eta):=\eta_2^{\perp}+iJ(\eta_2^{\perp});
\label{H-Bdef2}
\\
&=2g_j(\xi_2^{\perp},\eta_2^{\perp})+2i\,\omega(\xi_2,\eta_2^{\perp}).
\label{H-Bdef}
\end{align}
\end{subequations}
It is important not to confuse $\mathcal{F}^-(\eta)$ with  $\eta^-$.

\paragraph{Decomposition of $\Omega(X;\Rc)^j_{d_H,\Zh}$.}
As in \cite{Belov:2006jd} we next need to
 choose a complementary part of $V_2 \cap \Omega(X;\Rc)^j_{d_H,\Zh}$
inside $\Omega(X;\Rc)^j_{d_H,\Zh}$. The complication here is that
$\Omega(X;\Rc)^j_{d_H,\Zh}$ is \textit{not} a Lagrangian subspace.

We define a ``complementary part'' of this subgroup as follows.
The symplectic form $\omega_j$ on $V_{\Rh}$ defines a symplectic form
on the twisted cohomology $\Gamma=H(X;\Rc)^j_{d_H}$. This symplectic
form is integral valued on the image $\bar{\Gamma}$
 of the twisted $K$-group.
In turn, a choice of Lagrangian subspace $V_2\subset V_{\Rh}$
defines a Lagrangian subspace $\Gamma_2\subset \Gamma$. We denote by
$\bar{\Gamma}_2$ the corresponding lattice inside $\bar{\Gamma}$.
Now define $\Gamma_1$ to be an arbitrary complementary Lagrangian
subspace to $\Gamma_2$ such that the lattice $\bar{\Gamma}$
decomposes as $\bar{\Gamma}_1\oplus \bar{\Gamma}_2$. We now
\textit{define} the subspace $V_1\subset V_{\Rh}$ to consist of all
$d_H$-closed forms of total degree $j$ whose twisted cohomology
class lies in $\Gamma_1$:
\begin{equation}
V_1=\{R\in\Omega(X;\Rc)^j_{d_H}\,|\, [R]_{d_H}\in\Gamma_1\}.
\label{LdefV1}
\end{equation}
We denote the intersection $V_1\cap \Omega(X;\Rc)^j_{d_H,\Zh}$ by
$\bar{V}_1$,  i.e. the space  \eqref{LdefV1} with $\Gamma_1$ changed
to $\bar{\Gamma}_1$.

\begin{lemma}
$V_1$ defined by \eqref{LdefV1} is an isotropic subspace in $V_{\Rh}$.
\label{lem:1}
\end{lemma}

Note that $V_1$ and $V_2$ are \textit{not complementary}
subspaces. They have nonzero intersection $V_{12}:=V_1\cap V_2$ where
\begin{equation}
V_{12}=\{\text{$d_H$-exact forms in $V_2$}\}.
\end{equation}

\paragraph{Quadratic function $\Omega_0$.}
Recall that to write the cocycle $\varphi$ in the simple form
\eqref{varphi-final} we chose a $\Zh_2$-valued QRIF $\Omega_0$ with  vanishing
$\mu$ class. Such a choice of QRIF with $\mu=0$ is naturally
determined by   a Lagrangian decomposition of
$K^{j,h(\check{B})}(X)/\text{torsion}=\bar{\Gamma}_1\oplus
\bar{\Gamma}_2$ as follows: We define $\Omega_{\bar \Gamma_1 \oplus
\bar \Gamma_2}$ to be $=1$ on $\bar \Gamma_1$ and $\bar \Gamma_2$.
Its values on all other vectors is then determined by the
quadratic refinement law. More explicitly, any   $R\in\Omega(X;\Rc)^j_{d_H,\Zh}$ can be written
as $R=R_1+R_2$ where $R_1\in \bar{V}_1$ and $R_2\in
V_2\cap\Omega(X;\Rc)^j_{d_H,\Zh}$.  Now define
\begin{equation}
\Omega_{\Gamma_1\oplus \Gamma_2}(R):=e^{i\pi \omega_j(R_1,R_2)}.
\end{equation}
Since $V_1\cap V_2\ne\{0\}$ the
decomposition $R=R_1+R_2$ is not unique,
 but since  $R_1$ and $R_2$ are $d_H$-closed it
follows that $\Omega_{\bar\Gamma_1\oplus \bar\Gamma_2}(R)$
does not depend on a particular choice of decomposition.
Moreover $\Omega_{\Gamma_1\oplus\Gamma_2}$ takes values in $\{\pm 1\}$.
We choose $\Omega_{\Gamma_1\oplus\Gamma_2}$ as the QRIF $\Omega_0$ in \eqref{varphi-final}.

\paragraph{Partition function.}
Now one can solve equation \eqref{theta} via   Fourier
analysis. The expression for the partition function can be
summarized by the following theorems:

\begin{thm}
\label{thm:1}
The following Euclidean functional integral
\begin{multline}
\vartheta^{\eta}(a^+)
=\exp\Bigl[-\frac{\pi}{2}(H-B)(\eta^+,\eta^+)
+\frac{\pi}{2}B(a^+,a^+)-\pi(H-B)(a^+,\eta^+)\Bigr]
\\
\times\int_{\bar{V}_1/V_{12}} \mathscr{D} R
\,\exp\Bigl[-\frac{\pi}{2}(H-B)(R^+,R^+)
+\pi(H-B)(a^++\eta^+,R^+)\Bigr]
\label{Epath2}
\end{multline}
(where the integral goes over all closed forms $R\in \bar{V}_1$
modulo $d_H$-exact forms in $V_2$)
$a,\eta\in\Omega(X;\Rc)^j$
\begin{enumerate}
\item[a)]
satisfies the functional equation
\begin{equation}
\vartheta^{\eta}(a^++\lambda^+)=
\Omega_{\Gamma_1\oplus\Gamma_2}(\lambda)\,
e^{2\pi i \,\omega_j(\eta,\lambda)}\,
e^{\pi H(a^+,\lambda^+)+\frac{\pi}{2}H(\lambda^+,\lambda^+)}\,
\vartheta^{\eta}(a^+)
\end{equation}
for all $\lambda\in \Omega(X;\Rc)^j_{d_H,\Zh}$.
\item[b)] satisfies the equation
\begin{equation}
\vartheta^{\eta+\lambda}(a^+)=
\Omega_{\Gamma_1\oplus\Gamma_2}(\lambda)\,
e^{i\pi \,\omega_j(\eta,\lambda)}\,
\vartheta^{\eta}(a^+)
\end{equation}
\label{prop2}
\end{enumerate}
\end{thm}

If both $a$ and $\eta$ are twisted harmonic forms
(see Appendix B for the definition)
then one can evaluate the functional integral in \eqref{Epath2}
explicitly. The result is summarized by
\begin{thm}
\label{thm62} The functional integral \eqref{Epath2} for
$\eta,\,a\in\mathscr{H}(X;\Rc)^j_{d_H}$ is equal to
\begin{equation}
\theta^{\eta}(a^+)=\Nc_{V_1,V_2}(g)\,
\vartheta\left[\begin{smallmatrix} \eta_1
\\
\eta_2
\end{smallmatrix}\right](a^+)
\end{equation}
 where $\mathcal{N}_{V_1,V_2}(g)$ is a purely metric dependent factor coming from
the integration over the exact forms in \eqref{Epath2}, and
$\vartheta\left[\begin{smallmatrix}
\eta_1
\\
\eta_2
\end{smallmatrix}\right](a^+)
$ is the canonical theta function on the finite dimensional
torus $\mathscr{H}(X;\Rc)^j_{d_H}/\mathscr{H}(X;\Rc)^j_{d_H,\Zh}$
\begin{multline}
\vartheta\left[\begin{smallmatrix}
\eta_1
\\
\eta_2
\end{smallmatrix}\right](a^+)
=\exp\Bigl[-i\pi\omega_j(\eta_2,\eta_1)
+\frac{\pi}{2}B(a^+,a^+)-\pi(H-B)(a^+,\eta^+)\Bigr]
\\
\times \sum_{R\in\bar{\Gamma}_1^{h}-\eta_1}
\,\exp\Bigl[-\frac{\pi}{2}(H-B)(R^+,R^+)
+\pi(H-B)(a^++\eta^+_2,R^+)\Bigr].
\label{Epath3}
\end{multline}
Here $\eta=\eta_1+\eta_2$ according to the Lagrangian decomposition
of the space of harmonic forms
$\Gamma^h=\Gamma_1^h+\Gamma_2^h$.
\end{thm}

\begin{rem}
The form $(H-B)$ restricted to $V_1^+\times V_1^+$ is symmetric.
From \eqref{H-Bdef} it also follows that it vanishes on
$V_{12}$ and  $\Re (H-B)|_{V_1^+\times V_1^+}$ is positive
definite on the complement of $V_{12}$ inside $V_1$. In the theory
of theta functions the quadratic  form $(H-B)$ restricted to the
finite dimensional space $\Gamma_1^h:=V_1\cap
\mathscr{H}(X;\Rc)^{j}_{d_H}$
\begin{equation}
\tau(v_1^+) := \frac{i}{2} (H-B)(v_1^+,v_1^+)
\quad\text{for}\quad v_1\in\Gamma_1^h
 \label{cpx.period}
\end{equation}
is known as the complex period matrix.
\end{rem}

%

\begin{cor}
\label{cor61}
The partition function $\mathcal{Z}(a^+,a^-)$ is
\begin{equation}
\mathcal{Z}_{V_1,V_2,J}(a^+,a^-)
=e^{i\pi \Re\omega_j(\sigma(\check{A}_{\bullet}+\check{\mu}),
\mathcal{F}^-(\sigma(\check{A}_{\bullet}+\check{\mu})))}
e^{i\pi \omega_j(a^-,a^+)}\vartheta^{\sigma(\check{A}_{\bullet}+\check{\mu})}(a^+)
\label{Zfinal}
\end{equation}
where $\check{\mu}$ is defined in \eqref{varphi-final} and
determined by  a choice of Lagrangian subspaces $\Gamma_1$ and
$\Gamma_2$. The factor
$e^{i\pi\Re\omega_j(\sigma(\check{A}_{\bullet}+\check{\mu}),
\mathcal{F}^-(\sigma(\check{A}_{\bullet}+\check{\mu})))}$ is added
to ensure that the theta function does not depend on the choice of
Lagrangian subspace $\Gamma_1$. We stress that $\check{\mu}$ does
depend on the choice of Lagrangian decomposition $\Gamma_1\oplus
\Gamma_2$.
\end{cor}

%

\example{
The partition function \eqref{Zfinal} looks particularly simple on the real slice $a^-=(a^+)^*$
in the space of complexified gauge fields:
\begin{multline}
\mathcal{Z}_{V_1,V_2}(a;g)
=e^{i\pi \Re\omega_j(\sigma_{\bullet},
\mathcal{F}^-(\sigma_{\bullet}))
-i\pi\omega_j(a,\sigma_{\bullet})}
\\
\times
\int_{R\in \bar{V}_1/V_{12}}
\Ds R\,
\exp\Bigl[-i\pi\omega_j(R-\sigma,\mathcal{F}^-(R-\sigma))
-i\pi\omega_j(R,\sigma)
\Bigr]
\label{Z}
\end{multline}
where $\sigma=\sigma(\check{A}+\check{\mu})$ and
$\sigma_{\bullet}=\sigma(\check{A}_{\bullet}+\check{\mu})$.
}

\begin{rem}
\label{rem64}
We can use the base point $\check{A}_{\bullet}$ to
obtain a partition function coupled to the source $[\check{j}]$:
the partition function of the RR field coupled to the source $\check{j}$
is
\begin{equation}
\mathcal{Z}_{V_1,V_2}(a;\check{j}):=\mathcal{Z}_{V_1,V_2}(a)|_{\check{A}_{\bullet}=-\check{j}}.
\end{equation}
\end{rem}
This gives a physical interpretation to the choice of a basepoint
in our constructions above - it is equivalent to a choice of a background
RR current.

\paragraph{Quantum equation of motion.}
The infinitesimal version of the Gauss law \eqref{GSL2} for $R=d_Hc$ yields
a differential equation on $\mathcal{Z}$
\begin{equation}
\Bigl[D_{d_H c}+2\pi i\,\omega_{j+1}\bigl(c,F(\check{A})\bigr)\Bigr]\mathcal{Z}(a)=
0.
\label{qEOM}
\end{equation}
Now we can apply this equation to the partition function \eqref{Zfinal}
restricted to the real slice in the space of complexified gauge fields: $(a^+)^*=a^-$.
Taking into account that $\mathcal{Z}$ is a holomorphic section,
$D^-\mathcal{Z}=0$ we obtain the following

\begin{thm}[Quantum equation of motion]
The infinitesimal Gauss law yields the quantum equation of motion
\begin{equation}
d_H\bigl\langle \mathcal{F}^-(R-\sigma)\bigr\rangle_{\check{A},\check{j}}
=-F(\check{A})
\label{qEOM2}
\end{equation}
where $\mathcal{F}^-(v):=v_2^{\perp}+Jv_2^{\perp}$ for any
$v\in\Omega(X;\Rc)^j$, $\sigma=\sigma(\check{A}+\check{\mu})$ is
defined in \eqref{sigmadef}. $\langle
\Oc(R)\rangle_{\check{A},\check{j}}$ is the normalized correlation
function defined as the ratio of the Euclidean functional integral
\eqref{Epath2} with the insertion of $\Oc(R)$ and the same integral
without the insertion.
\end{thm}

\begin{proof}
{}From \eqref{H-Bdef2} it follows that
$(H-B)(v^+,w^+)=2i\,\omega(v,\,\mathcal{F}^-(w))$. A straightforward
calculation yields
\begin{equation*}
\frac{1}{2\pi i\, \mathcal{Z}}\,D^+\mathcal{Z}(a^+,a^-)
=\omega_j(\delta a,\, \bigl\langle \mathcal{F}^-(R
-\sigma)\bigr\rangle_{\check{A}})+\omega_j(\delta a^+,(a^+)^*-a^-).
\end{equation*}
Now if we restrict it to the real slice $(a^+)^*=a^-$ the last term disappears.
To obtain \eqref{qEOM2} one needs to substitute $\delta a=d_H \alpha$ into
the equation above, integrate by parts and compare with \eqref{qEOM}.
\end{proof}

\clearpage
\section{$B$-field dependence}
\label{sec:Bfield}
\setcounter{equation}{0}
In the previous sections we considered the partition function $\mathcal{Z}_{\check{B}}$
as a function of the $B$-field for $\check{B}\in\check{Z}^2(X)$.
However one expects that the partition function should
depend only on the gauge equivalence class of the $B$-field.
In other words, the partition function must descend to
 a function over a connected component of $\check{H}^3(X)$.

\begin{figure}[!h]
\centering
\includegraphics[width=260pt]{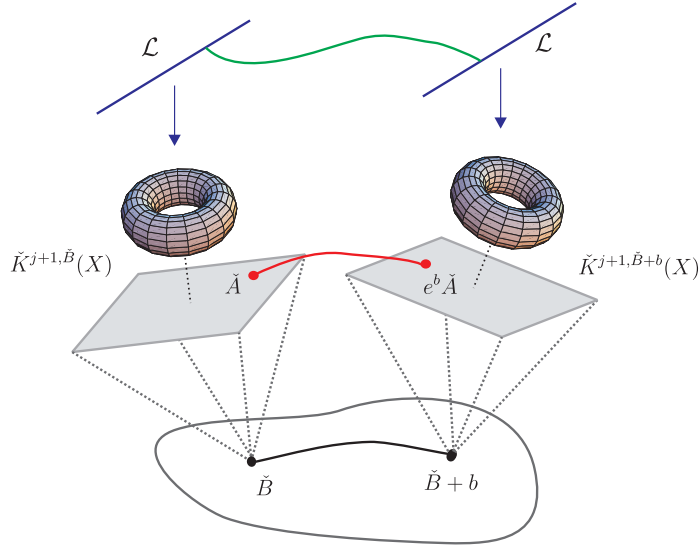}
\caption{The space of D-brane currents is fibered
over the space $\check{Z}^2(X)$ of $B$-fields. On each connected component
of $\check{Z}^2(B)$ there is a path connecting two points,  say $\check{B}$
and $\check{B}+b$. There is a natural lift of this path into the total
space: $(\check{B},\check{A})\mapsto (\check{B}+b,e^b\check{A})$.
In addition we have a Chern-Simons line bundle over the space of
D-brane currents. We now have to lift the path on $\check{Z}^2(X)$
to the total space of the line bundle such that it maps covariantly
constant sections to covariantly constant sections.
}
\label{fig:b-fibration}
\end{figure}

The partition function \eqref{Z} is manifestly invariant under the $B$-field gauge transformation:
all terms in the partition function are constructed from the objects which
are invariant under the $B$-field gauge transformation.

The $B$-field dependence of the partition function \eqref{Z}
can be summarized by the following variational formula:
\begin{multline}
\frac{1}{2\pi i}\,\delta_b\log\mathcal{Z}_{V_1,V_2}(a)
=-\frac12\,\Bigl\langle
\omega_j\bigl(u^{-1}\delta b\wedge \mathcal{F}^-(R-\sigma),
\mathcal{F}^-(R-\sigma)\bigr)
\\
-\omega_j\bigl(u^{-1}\delta b\wedge (\sigma_{\bullet})_2^{\perp},
(\sigma_{\bullet})_2^{\perp}\bigr)
\Bigr\rangle
\label{dbSL}
\end{multline}
In deriving \eqref{dbSL} we used the following
\begin{lemma}
\label{lem:81} When the $B$-field changes by $\check{B}\mapsto
\check{B}+\delta b$ where $\delta b\in \Omega^2(X)$ and  elements of
$\Omega(X;\Rc)^j$ are changed by $\delta_b v=u^{-1}\delta b\wedge
v$. The bilinear form $\omega_j(v,\mathcal{F}^-(w))$ changes by
\begin{equation}
\delta_b\omega_j(v,\mathcal{F}^-(w))=\omega_j(u^{-1}\delta b\wedge
\mathcal{F}^-(v),\mathcal{F}^-(w)).
\label{Bvar}
\end{equation}
The right hand side is a symmetric $\Rh$-bilinear form.
\end{lemma}
\begin{proof}
The statement of lemma follows from the fact that the Lagrangian subspace $V_2$
is invariant under $b$-transformations $e^{u^{-1}b}V_2\subseteq V_2$ where $b\in \Omega^2(X)$.
From this invariance and equation
\begin{equation}
\delta_b v=u^{-1} \delta b\wedge v_2
+u^{-1} \delta b\wedge v_2^{\perp}
=(\delta_b v)_2+(\delta_b v)_2^{\perp}
\label{lm1}
\end{equation}
it follows that
\begin{equation}
(\delta_b v)_2^{\perp}-u^{-1} \delta b\wedge v_2^{\perp}
=u^{-1} \delta b\wedge v_2-(\delta_b v)_2\in V_2.
\label{lm2}
\end{equation}
This equation will allow us to substitute $u^{-1} \delta b\wedge v_2^{\perp}$
instead of $(\delta_b v)_2^{\perp}$ in several terms below.
The variation is
\begin{multline*}
\delta_b\omega_j(v,\mathcal{F}^-(w))=
\omega_j((\delta_bv)_2,w_2^{\perp})+\omega_j((\delta_bv)_2^{\perp}, iJw_2^{\perp})
+\omega_j(v_2,(\delta_bw)_2^{\perp})+\omega_j(v_2^{\perp},iJ(\delta_b w)_2^{\perp})
\\
\stackrel{\ref{lm1},\,\ref{lm2}}{=}
\omega_j(u^{-1}\delta b\wedge v_2+u^{-1}\delta b\wedge v_2^{\perp},w_2^{\perp})
+\omega_j(u^{-1}\delta b\wedge v_2^{\perp},iJw_2^{\perp})
\\
+\omega_j(v_2,u^{-1}\delta b\wedge w_2^{\perp})
+\omega_j(v_2^{\perp},iJ(u^{-1}\delta b\wedge w_2^{\perp})).
\end{multline*}
Now from the invariance of the symplectic form under $b$-transformations
\eqref{btrinv} and
$0=\omega_j(u^{-1}\delta b\wedge Jv_2^{\perp},Jw_2^{\perp})$ it follows
that the equation above can be rewritten as in \eqref{Bvar}.
\end{proof}

\subsection{Type IIA partition function}
It is well known that the equation of motion for the $B$-field
contains an extra metric dependent term \cite{Vafa:1995fj,Duff:1995wd,Sethi:1996es}:
\begin{equation}
\ell_s^{-4}d(e^{-2\phi}*H)
=\ell_s^{-6}R_0\wedge *R_2+
\ell_s^{-2}R_2\wedge *R_4-\frac12\,R_4\wedge R_4
+X_8(g)
\end{equation}
where $X_8(g)=\frac{1}{48}(p_2-\lambda^2)$ and $\lambda=-\frac{p_1}{2}$
and $p_i(g)$ denotes the standard representative of the Pontrjagin class
in terms of traces of curvatures.
This term naturally comes from the reduction of $M$-theory.
It is clear that the action of type IIA must contain some $B$-field
dependent term whose variation yields $X_8(g)$. In the literature this term is
usually taken to be
\begin{equation} e^{-2\pi i \int_M B\wedge X_8}
\end{equation}
but this is incorrect because $X_8$ does
not have integral periods. Thus, this term in the action is not
 gauge invariant under
  large $B$-field gauge transformations. Moreover, it is not well-defined in
  the presence of topologically nontrivial $B$-fields.
In this subsection we will show how our careful formulation of the
partition function \eqref{Z} cures this problem.

The partition function \eqref{Z} is invariant under   $B$-field
gauge transformations. Thus   we can only multiply it by a  gauge invariant quantity.
Although $X_8$ does not have integral periods it was shown
in \cite{Diaconescu:2000wy} that
\begin{equation}
\frac{\lambda^2}{8}-X_8=30\hat{A}_8
\end{equation}
has an integral lift $\Theta \in H^8(X,\Zh)$ whose image in de Rham
cohomology is  $30\hat{A}_8$. Therefore,
   the term $e^{2\pi i\int_M B\wedge 30\hat{A}_8}$ is invariant under
   large $B$-field gauge transformations and can be defined in the presence
   of topologically nontrivial $B$-fields. Therefore, we can safely multiply our partition
   function by this term.   The question is whether
we get the term $-\int_M \delta B\wedge \lambda^2/8$ by varying the
RR partition function \eqref{Z}.

Suppose that there is no external current and the characteristic
class $\mu$ vanishes. Then we can choose a base point $\check{A}_{\bullet}=0$.
So $\varepsilon:=\sigma(\check{\mu})$ is a globally well defined
form of total degree $j$ such that $2\,\varepsilon\in \Omega(X;\Rc)^j_{d_H,\Zh}$.
The variation \eqref{dbSL} of the partition function under the change
of the $B$-field is
\begin{equation*}
\frac{1}{2\pi i}\,\delta_b\log\mathcal{Z}_{V_1,V_2}(0)
=-\frac12\,\Bigl\langle
\omega_j\bigl(u^{-1}\delta b\wedge \mathcal{F}^-(R-\varepsilon),
\mathcal{F}^-(R-\varepsilon)\bigr)
-\omega_j\bigl(u^{-1}\delta b\wedge (\varepsilon)_2^{\perp},
(\varepsilon)_2^{\perp}\bigr)
\Bigr\rangle.
\end{equation*}
The characteristic $\varepsilon$ can be decomposed as $\varepsilon=
\varepsilon_1+\varepsilon_2$.
One sees that the variation of the partition function indeed contains
an extra term: $\frac12\,(\varepsilon_1)_{2}^{\perp}
\wedge \phi_0((\varepsilon_1)_2^{\perp})$.
In general, the projection $(\varepsilon_1)_2^{\perp}$ of the characteristic $\varepsilon_1$ depends
on the choice of  Lagrangian subspace $V_2$.
In type IIA theory there is a natural choice of the Lagrangian
subspace $V_2$ given by the forms of degree $6,\,8$ and $10$
(see \eqref{exV2} for more details).
Reference \cite{Diaconescu:2000wy}
calculated the characteristic $\varepsilon_1$ (or more precisely its projection
$(\varepsilon_1)_2^{\perp}$) for   vanishing
$B$-field as $\varepsilon_1=\frac{\lambda}{2}+\dots$.
One expects that for topologically trivial $B$-field
$\varepsilon_1=e^b(\frac{\lambda}{2}+\dots)$ where ``$\dots$'' denote higher
degree terms. Thus for topologically trivial $B$-field the variation of
\begin{equation}
e^{-2\pi i \int_M b\wedge 30\hat{A}_8}
e^{i\pi \Re \omega(\varepsilon_1,\mathcal{F}^-(\varepsilon_1))}
\end{equation}
is exactly $+2\pi i \int_M \delta b\wedge X_8$.

Now let us try to generalize these arguments for a
topologically nontrivial $B$-field.
Unfortunately we do not know how to calculate $\varepsilon_1$ directly,
so we conjecture its form, but our conjecture is strongly supported by
the known equation of motion of the $B$-field, which should apply in the
topologically nontrivial case as well.
It is clear that to be able to use the arguments above $\varepsilon_1$
must be related to $\lambda/2$. On the other hand  $\varepsilon_1$
must be a $d_H$-closed form. Therefore we essentially want
 $\lambda$ to have a lift to the $d_H$-cohomology.
However this condition cannot be taken literally, and indeed it is
easy to make backgrounds which are obviously consistent for which
$[\lambda \wedge H]_{dR}$ is nonzero. (Take the product of a WZW
model with a CY space.)
 Recall that on a spin manifold $\lambda$ is
the preferred  integral lift of the fourth Wu class $\nu_4\in H^4(X;\Zh/2)$.
Thus an arbitrary integral lift of the Wu class $\nu_4$ is $[\lambda+2\rho]$ where $\rho$
is  a closed $4$-form with integral periods.
We now require that there exists some $\rho$ such $\lambda+2\rho$ has
a lift to the $d_H$-cohomology. This in means
that $[(\lambda+2\rho)\wedge H]_{dR}=0$ and $\{H,H,\lambda+2\rho\}=0$
where $\{\cdot,\cdot,\cdot\}$ denotes the Massey product.
If such a $\rho$ exists we conjecture that $\varepsilon_1=(\frac12\lambda+\rho)+\dots$.
In this case the variation of the phase
\begin{equation}
e^{-2\pi i \int_{Y}H\wedge (30\hat{A}_8-\frac12 \rho(\rho+\lambda))}
e^{i\pi \Re \omega(\varepsilon_1,\mathcal{F}^-(\varepsilon_1))}
\end{equation}
indeed yields exactly $+2\pi i \int_M \delta b\wedge X_8$.
Here $Y$ is an $11$-manifold which bounds the $10$-dim manifold $X$.
Notice that the first exponential does not depend on the extension
since both $30\hat{A}_8$ and $\frac12\,\rho(\rho+\lambda)$ have
integral periods.

We regard the above argument as strong evidence that there is a new
topological restriction on
  consistent backgrounds of string theory, namely,
the fourth Wu class $\nu_4$ must have a lift $\lambda + 2 \rho$
whose projection modulo torsion is the first term in a cocycle for
$d_H$-cohomology.

The above condition would be compatible with the following natural
condition on IIA backgrounds: $\nu_4$ must have an integral lift
which itself has a lift to twisted $K$-theory. The first condition
for such a lift would be:
\begin{equation}\label{morelikely}
(Sq^3 + h) (\lambda + 2 \rho) =0
\end{equation}
We speculate that this is in fact a topological restriction on IIA
backgrounds.  Unlike the condition we have derived, which is a
consequence of \eqref{morelikely}, this condition is cleanly stated
at the level of integral cohomology. For $h=0$ this condition was in
fact derived carefully in \cite{Diaconescu:2000wy}. A generalization
to nonzero $h$ was derived in \cite{Diaconescu:2003bm}.
Unfortunately, our equation \eqref{morelikely} is not quite
compatible with equation $(11.10)$ of \cite{Diaconescu:2003bm}, and
we  suspect that the  equation in  \cite{Diaconescu:2003bm} should
be corrected to \eqref{morelikely}. This point obviously deserves
further study, but is outside the scope of the present paper.

\clearpage
\section{Action and equations of motion}
\label{sec:action}\setcounter{equation}{0}
The action for the self-dual field is essentially
the complex period matrix \eqref{cpx.period} extended
from the twisted K-group to the space of $d_H$-closed forms.
The purpose of this section is
to describe this extension in detail.

\subsection{Classical action}
First we need to extend the definition of the complex period matrix
\eqref{cpx.period} defined on the cohomology to the infinite
dimensional  symplectic  vector space $V_{\Rh}=\Omega(X;\Rc)^j$
consisting of forms of total degree $j$.

\paragraph{$(X,g_E)=$ Riemannian manifold.}
Following the  discussion in section~\ref{subsec:part} we choose an
orthogonal coordinate system on $V_{\Rh}$ to be $V_2\oplus
V_2^{\perp}$ where $V_2$ is a Lagrangian subspace and
$V_2^{\perp}=J(V_2)$ is its orthogonal complement with respect to
the Hodge metric. From the positivity of $g_E$ it follows that
$V_2\cap V_2^{\perp}=\{0\}$. Recall that the Hodge complex structure
is compatible with the symplectic
\begin{wrapfigure}{l}{150pt}
\includegraphics[width=145pt]{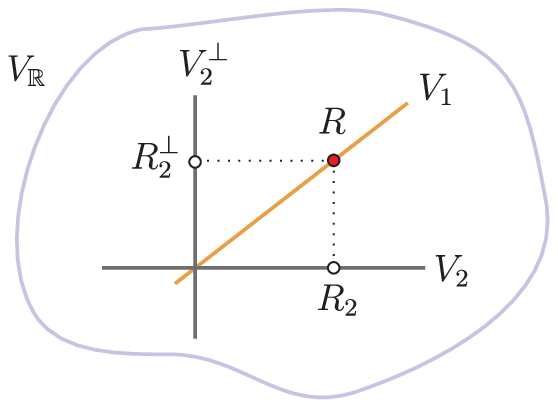}
\end{wrapfigure}
structure, thus
$V_{\Rh}=V_2\oplus V_2^{\perp}$ is a
Lagrangian decomposition.
So any form $v\in V_{\Rh}$ can be uniquely
written in the form
$v=v_2+v_2^{\perp}$ for some $v_2\in V_2$ and $v_2^{\perp}\in V_2^{\perp}$.

Let $V_1$ be another Lagrangian subspace. A choice of Lagrangian
decomposition $\Gamma=\Gamma_1\oplus \Gamma_2$ of the twisted cohomology
$\Gamma=H(X;\Rc)^j_{d_H}$ defines a \textit{canonical} choice of
$V_1$. However we postpone this discussion till the next paragraph.
Now any element $R$ from the Lagrangian subspace $V_1$ can be
written as
\begin{equation}
R=R_2+R_2^{\perp}
\label{RRperp_dec}
\end{equation}
where $R_2$ and $R_2^{\perp}$ are not independent but related by some linear function
(see the figure). From \eqref{H-Bdef} it follows
that the Euclidean action is
\begin{equation}
S_E(R^+):=i\pi\omega_j(R,\mathcal{F}^-(R))
\quad\text{where}\quad
\mathcal{F}^-(R):=R_2^{\perp}+iJ R_2^{\perp}.
\label{Eaction}
\end{equation}

\paragraph{$(M,g)=$ Lorentzian manifold.}
The action in  Lorentzian signature can obtained
from \eqref{Eaction} by Wick rotation:
\begin{equation}
S_L(R):=\pi\omega_j(R, \mathcal{F}^+(R))
\quad\text{where}\quad
\mathcal{F}^+(R):=R_2^{\perp}+I R_2^{\perp}.
\label{Maction}
\end{equation}
This action depends on the choice of Lagrangian subspace $V_2$. For a
Riemannian manifold a choice of $V_2$ automatically defines the
Lagrangian decomposition $V_{\Rh}=V_2\oplus  J(V_2)$. For a
Lorentzian manifold this is not true, and we need \textit{to
constrain} the choice of $V_2$ by the requirement
\begin{equation}
V_2\cap I(V_2) =\{0\}.
\label{V2constr}
\end{equation}
This condition means that $V_2$ should not contain
self- or anti self-dual forms.

In principle, $V_2$ can be an arbitrary Lagrangian subspace
satisfying the constraint \eqref{V2constr}. Recall that on any Lorentzian
manifold $M$ there exists a nowhere vanishing
timelike vector field $\xi$. It can be
used to define a Lagrangian subspace
$V_2(\xi)\subset \Omega(M;\Rc)^j$ via
\begin{equation*}
V_2(\xi):=\{\omega\in \Omega(M;\Rc)^j\,|\,i_{\xi}\omega=0\}.
\end{equation*}

\paragraph{How to choose the Lagrangian subspace $V_1$.}
Again we follow the discussion of section~\ref{subsec:part}. The
symplectic form $\omega_j$ on $V_{\Rh}$ defines a symplectic form on
the twisted cohomology $\Gamma=H(M;\Rc)^j_{d_H}$. In turn, a choice
of Lagrangian subspace $V_2\subset V_{\Rh}$ defines a Lagrangian
subspace $\Gamma_2\subset \Gamma$. We choose $\Gamma_1$ to be an
arbitrary complementary Lagrangian subspace, so
$\Gamma=\Gamma_1\oplus \Gamma_2$. Using this data we define the
subspace $V_1\subset V_{\Rh}$ to consist of all $d_H$-closed forms
of total degree $j$ whose cohomology class lies in $\Gamma_1$:
\begin{equation}
V_1 :=\{R\in\Omega(M;\Rc)_{d_H}^j\,|\, [R]_{d_H}\in\Gamma_1\}.
\label{defV1}
\end{equation}
$V_1$ defined by \eqref{defV1} is an isotropic subspace of
$\Omega(M;\Rc)^j$. Note that the  Lagrangian subspace $V_1$ really
depends on the choice of $B$-field:   If $b\in\Omega^2(X)$
then the two subspaces must be related by
\begin{equation}
V_1\vert_{\check{B}+b}=e^bV_1\vert_{\check{B}}.
\end{equation}

\paragraph{Equations of motion.} The variational problem for the action \eqref{Maction}
is summarized by the following theorem:
\begin{thm}
\label{thm:71}
Let $V_1\subset V_{\Rh}$ be a Lagrangian subspace defined by
\eqref{defV1}, and let $R\in V_1$ be a $d_H$-closed form. Then the action
\begin{equation}
S_L(R)=\pi\omega_j(R, \mathcal{F}^+(R))
=-\pi g_j(R_2^{\perp},R_2^{\perp})+\pi\omega_j(R_2,R_2^{\perp})
\label{SMaction}
\end{equation}
where $\mathcal{F}^+(R):=(R)_2^{\perp}+I (R)_2^{\perp}$
has the following properties:
\begin{subequations}
\begin{enumerate}
\item[a.] Variation with respect to $R\mapsto R+d_H\delta c$
where $\delta c\in\Omega(M;\Rc)^{j-1}_{cpt}$ is
\begin{equation}
\delta S_L(R)=2\pi\omega_{j+1}( u\delta c
,d_H\mathcal{F}^+(R)).
\label{SDflux}
\end{equation}

\item[b.] Stationary points of the action are the solutions of the equation:
\begin{equation}
d_H\mathcal{F}^{+}(R)=0.
\label{decF-}
\end{equation}
\end{enumerate}
\end{subequations}
\end{thm}

\begin{thm}
An arbitrary $d_H$-\textbf{closed} self-dual form $\mathcal{F}^+$ can
be written in form $\mathcal{F}^+(R)$ for some $R\in V_1$.
\end{thm}
The proof is similar to the one presented in \cite{Belov:2006jd}.

\paragraph{Gauge symmetries.} The discussion here is identical to that
in   \cite{Belov:2006jd}. Let
\begin{equation}
V_{12}:=V_1\cap V_2=\{d_H\text{-exact forms in }V_2\}.
\end{equation}
Then we have the
\begin{thm} The action \eqref{SMaction} has two types of gauge symmetries:
\begin{enumerate}
\item[a.] It manifestly invariant with respect $\check{C}\mapsto \check{C}+\omega$
where $\omega\in \Omega(M;\Rc)^{j-1}_{d_H,\Zh}$.
\item[b.] It is invariant under a shift $R\mapsto R+v_{12}$ where $v_{12}\in (V_1\cap V_2)_{cpt}$:
\begin{equation}
S_L(R+v_{12})=S_L(R).
\label{extragauge}
\end{equation}
Moreover the self-dual field $\mathcal{F}^+$  does not depend on $v_{12}$:
\begin{equation}
\mathcal{F}^+(R+v_{12})=\mathcal{F}^+(R).
\end{equation}
\end{enumerate}
\end{thm}

{}From this theorem it follows that
the gauge symmetry $R\mapsto R+v_{12}$ does not affect classical equations.
However this extra gauge symmetry has to be taken into account
in the quantum theory.

\paragraph{Coupling to the sources.}
In section~\ref{sec:defCSAB}  we mentioned that a $D$-brane configuration
defines a twisted differential character, the RR current
$[\check{j}]\in\check{K}^{j+1,\check{B}}(M)$. As has been
emphasized by D. Freed, the   Ramond-Ramond field
should be viewed as a trivialization of the total RR current
$[\check j + \check \mu]$.
This trivialization only exists   if $[\check{j}]$ satisfies the tadpole
constraint:
\begin{equation}
x([\check{j}])=\mu
\end{equation}
where $\mu$ is the torsion class appearing in the definition of the
Chern-Simons functional. The curvature $j_D:=F(\check{j})$ of the
differential cocycle $\check{j}$ is what is  usually called RR
current. In \cite{Green:1996dd,Cheung:1997az,Minasian:1997mm} a
formula for the RR current was derived:
\begin{equation}
j_D=\sqrt{\hat{A}}\ch_{\check{B}}(f_![P])\,\delta(W)
\end{equation}
Here $W$ is the worldvolume of the D-brane and  $\delta(W)$ is a
$\delta$-form representative of the class Poincar\'{e} dual to
$[W]$,  and  $f:W\hookrightarrow M$. $\hat{A}$ is differential form
representing the A-roof genus, $P$ is a twisted bundle on $W$, and
$\ch_{\check{B}}(f_![P])$ is a form representing the twisted Chern
character of the pushforward of the twisted bundle $[P]$ (in
particular, it depends on the gauge field on the $D$-brane).

If the tadpole constraint is satisfied then $\check{\mu}-\check{j}$
is a topologically trivial character. In section~\ref{sec:partfunc}
we denoted its trivialization by $\sigma_{\bullet}$:
\begin{equation}
d_H\sigma_{\bullet}=-j_D.
\end{equation}
From Remark~\ref{rem64} one obtains the following
\begin{thm}
\label{thm:91}
The Lorentzian action in the presence of $D$-branes is
\begin{subequations}
\begin{equation}
S_L(R)=\pi\omega_j(R-\sigma_{\bullet},\mathcal{F}^+(R-\sigma_{\bullet}))
+\pi\omega_j(R,\sigma_{\bullet})
-\pi\omega_j((\sigma_{\bullet})_2,(\sigma_{\bullet})_2^{\perp})
\label{SLD}
\end{equation}
for $R\in \bar{V}_1$.
The variation of the action with respect to $\delta R=d_H\delta c$ is given by
\begin{equation}
\delta S_L(R)=2\pi\,\omega_{j+1}(u\delta c,\,d_H\mathcal{F}^+(R-\sigma_{\bullet})
-j_D).
\end{equation}
The variation of the action under change of the $B$-field is
\begin{equation}
\delta_b S_L(R)=\pi\omega_j(u^{-1}\delta b\wedge \mathcal{F}^+(R-\sigma_{\bullet}),
\mathcal{F}^+(R-\sigma_{\bullet}))
-\pi\omega_j(u^{-1}\delta b\wedge (\sigma_{\bullet})_{2}^{\perp},
(\sigma_{\bullet})_2^{\perp}).
\end{equation}
\end{subequations}
\end{thm}

\example{\noindent The action can also be written in the form
\begin{equation}
S_L(R)=-\pi
g_j((R-\sigma_{\bullet})_2^{\perp},(R-\sigma_{\bullet})_2^{\perp})
+\pi
\omega_j((R)_2,(R)_2^{\perp})+2\pi\,\omega_j((R)_2^{\perp},(\sigma_{\bullet})_2)
\label{SLD2}
\end{equation}
The first term represents the kinetic terms for RR fields.  The
second term is the analog of the Chern-Simons interaction. Note that
it does not depend on the source $\sigma_{\bullet}$. The third term
is the usual ``electric'' coupling to the sources. }

Denoting $G_s = -\sigma_\bullet$ and $G=R-\sigma_{\bullet}$ we
obtain the action written informally in the introduction. Of course,
for localized sources such as branes, the expressions quadratic in
$\sigma_{\bullet}$ will require regularization. This is to be
expected when taking into account backreaction in supergravity, and
renormalization of these terms relies on an ultraviolet extension of
the supergravity theory.

It is interesting to take special note of the ``anomalous
couplings'' of the source to the RR current. This is represented by
the second term in \eqref{SLD} and by the third term in
\eqref{SLD2}. In the past literature on D-brane couplings there has
been some confusion about a crucial factor of two in this coupling.
In references \cite{Green:1996dd,Minasian:1997mm} the anomalous
coupling was written with a factor of two too large and there was a
compensating mistake in a reality condition. In references
\cite{Cheung:1997az,Freed:2000tt} this factor of two was corrected,
and particularly stressed in the second reference,  but without a
complete explanation of how to handle the self-duality of the RR
fields. It is very nice that the correct factor of two appears
automatically in our present formalism. Moreover, note that when we
rearrange terms so that there is only an electric coupling, as in
\eqref{SLD2} the strength of the coupling is two times as large.

\subsection{Examples}
\label{sec:examples}
One of the main results of this paper is that
the action for RR fields is \textit{not} unique:
the actions are parameterized by Lagrangian subspaces
$V_2$ and $\Gamma_1$. In this section we illustrate
this result by 4 examples:
\begin{itemize}
\item The action for RR fields of type IIA: 1) on an arbitrary
Lorentzian $10$-manifold; 2) on a product space $\Rh\times N$
where $N$ is a compact $9$-manifold.
\item The action for RR fields of type IIB:
on a product space $\Rh\times N$ for different choices
of $V_2$.
\end{itemize}

\subsubsection{RR fields of type IIA: Example 1}
The symplectic vector space $V_{\Rh}=\Omega(M;\Rc)^0$
can be written as:
\begin{equation}
\Omega(M;\Rc)^0=\underbrace{\Omega^0\oplus u^{-1}\Omega^2\oplus
u^{-2}\Omega^{4}}_{I(V_2)}\oplus \underbrace{u^{-3}\Omega^6\oplus u^{-4}\Omega^8
\oplus u^{-5}\Omega^{10}}_{V_2}
\label{exV2}
\end{equation}
It contains a natural Lagrangian subspace $V_2$ which is spanned by
forms of degree $6,\,8,\,10$ and is invariant under the $b$-transform: $e^{u^{-1}b}V_2\subseteq V_2$
where $b\in \Omega^2(M)$.
The Lagrangian subspace $V_2$ defines a Lagrangian subspace
$\Gamma_2$ in the even twisted cohomology $H(M;\Rc)^0_{d_H}$.
Let $\Gamma_1$ be an arbitrary complementary Lagrangian subspace. $V_1$ is defined by
\begin{equation}
V_1=\{R\in\Omega(M;\Rc)^0_{d_H}\,|\, [R]_{d_H}\in \Gamma_1\}.
\end{equation}

$R\in V_1$ decomposes as
\begin{equation}
R=\underbrace{R_0+u^{-1}R_2+u^{-2}R_4}_{R_2^{\perp}}
+\underbrace{u^{-3}R_6+u^{-4}R_8+u^{-5}R_{10}}_{R_2}.
\label{RRperpIIA}
\end{equation}
Let us stress once again that cohomology classes of $R_6,R_8,R_{10}$ are
\textit{functions} of $R_0,R_2,R_4$.

Substituting \eqref{RRperpIIA} into \eqref{SMaction} one finds
that the self-dual flux $\mathcal{F}^+(G)$ is
\begin{equation}
\mathcal{F}^+(G)=
G_0+u^{-1}G_2+u^{-2}G_4
+\ell_s^{-2}  u^{-3}\,*G_4
-\ell_s^{-6}  u^{-4}\,*G_2
+\ell_s^{-10}u^{-5}\,*G_0
\label{SDex1}
\end{equation}
where $G=R-\varepsilon$. Recall that  $\varepsilon = \sigma([\check \mu])$   in the absence of
external current.
The Lorentzian action is
\begin{multline}
e^{iS_{\texttt{IIA}}(R)}
=\exp\left[-\frac{i\pi}{\ell_s^{10}}\int_M G_0\wedge *G_0
-\frac{i\pi}{\ell_s^6}\int_M G_2\wedge *G_2
-\frac{i\pi}{\ell_s^2}\int_M G_4\wedge *G_4
\right.
\\
\left.
-i\pi\int_M(R_{10}\wedge R_0-R_8\wedge R_2+R_6\wedge R_4)
-2\pi\int_M(R_0\wedge\varepsilon_{10}-R_2\wedge\varepsilon_{8}
+R_4\wedge\varepsilon_{6})
\right]
.
\label{IIAex1}
\end{multline}
Note that in the first line the shifted curvature $G$ appears while
the second line is written using only $R$.
One might suspect that this action functional depends on extra
degrees of freedom contained in $5$-, $7$- and $9$-form gauge
potentials, but this is not the case because of the definitions of
fields in terms of Lagrangian subspaces.

The variation of the action \eqref{IIAex1} with
respect to $R\mapsto R+d_H\delta c$ yields
the equation $d_H\mathcal{F}^+(G)=0$ which in components yields
the Bianchi identities:
\begin{subequations}
\begin{align}
&dG_0=0,\quad dG_2-H\wedge G_0=0,\quad
dG_4-H\wedge G_2=0;
\\
\intertext{together with the equations of motion}
&d(\ell_s^{-2}*G_4)-H\wedge G_4=0,
\quad
d(\ell_s^{-4}*G_2)+H\wedge *G_4=0.
\end{align}
\end{subequations}

The kinetic terms in the action \eqref{IIAex1} coincide with the
kinetic terms of the RR fields in the string frame. The topological
term is different from what   is usually written. The usual
discussion proceeds as follows: Suppose that the $10$-dimensional
manifold is a boundary of an $11$-manifold $Y$ and that $H$ and
$R_4$ admit extensions  $\tilde H$ and $\tilde R_4$ to $Y$ . In this
situation the Chern-Simons term is usually defined by
$e^{-i\pi\int_Y \tilde{H}\tilde{R}_4\tilde{R}_4}$. In order for this
term to be well defined the integral
$\oint_{Y}\tilde{H}\tilde{R}_4\tilde{R}_4$ over an arbitrary closed
$11$-manifold would have to be an even integer. Unfortunately, in
general, such extensions need not exist; even when they do exist
they are not correctly quantized, so that the expression is not
well-defined. Moreover, if the extension is just defined as an
extension of differential forms the expression varies continuously
with a choice of extension.
 Our expression   \eqref{IIAex1} nicely resolves all these problems.
 To make contact with the usual expression, suppose
 that $M$ is a boundary of an $11$ manifold $Y$ and the
 differential $K$-character $[\check{C}]$ extends to a character $[\check{\tilde{C}}]$
 over $Y$. Then, in particular, $R_{2p},H$ admit extensions $\tilde R_{2p}, \tilde H$
 to $Y$ satisfying the Bianchi identity. In this case
\begin{equation}
e^{-i\pi\int_{\pd Y} (R_{10}\wedge R_0-R_8\wedge R_2+R_6\wedge R_4)}
=e^{-i\pi \int_Y d(\tilde{R}_{10}\wedge \tilde{R}_0-\tilde{R}_8
\wedge \tilde{R}_2+\tilde{R}_6\wedge \tilde{R}_4)}
=e^{-i\pi\int_Y \tilde{H}\wedge \tilde{R}_4\wedge \tilde{R}_4}
\end{equation}
where in the last equality we used the Bianchi identity.
Using this one sees that the local expression in \eqref{IIAex1} has the
same local variation as the standard CS term. There is no dependence on the
choice of extension since $\tilde{H}\tilde{R}_4\tilde{R}_4$
is \textit{an exact} form.

\begin{rem}
 Suppose we change the $B$-field from $\check{B}$
to $\check{B}+b$. The curvature of the  $RR$ field also changes:
$R(\check{B}+b)=e^{u^{-1}b}R(\check{B})$. The  ``Chern-Simons'' for
$\check{B}+b$ is related to that for $\check{B}$ by the
multiplicative factor
\begin{equation}
\exp\biggl[
-i\pi\int_M\Bigl[
b\, R_4R_4+b^2\,R_2R_4
+\frac13\,b^3\,(R_2^2+R_0R_4)
+\frac14\,b^4\,R_0R_2
+\frac{1}{20}\,b^5\,R_0^2
\Bigr]
\biggr]
\end{equation}
where $R$ is the curvature of the RR field for
the $B$-field $\check{B}$.
\end{rem}

\paragraph{IIA action in the presence of D-branes.}
It is also important to stress that our expression makes sense in
the presence of $D$-brane sources.  In the presence of sources
 the standard CS term
is not well defined since $\tilde{H}\tilde{R}_4\tilde{R}_4$
is not even a closed form. Nevertheless, \eqref{IIAex1} remains well defined.

The D-brane current $j_D$ can be written as
$j_D=j_1+u^{-1}j_2+\dots+u^{-4}j_9$. Let $-\sigma$
be its trivialization: $j_D=-d_H\sigma$.
Substituting the expansion \eqref{RRperpIIA} into \eqref{SLD2}
one finds the Lorentzian action
\begin{multline}
S_{\texttt{IIA}}(R)
=-\frac{\pi}{\ell_s^{10}}\int_M G_0\wedge *G_0
-\frac{\pi}{\ell_s^6}\int_M G_2\wedge *G_2
-\frac{\pi}{\ell_s^2}\int_M G_4\wedge *G_4
\\
-\pi\int_M(R_{10}\wedge R_0-R_8\wedge R_2+R_6\wedge R_4)
-2\pi\int_M(R_0\wedge \sigma_{10}-R_2\wedge \sigma_8+R_4\wedge\sigma_6)
\label{IIADex1}
\end{multline}
where $G=R-\sigma$.
The first three terms are the standard couplings to the magnetic branes,
while   the last term is the WZ term for the D-brane.
Under the change $R\mapsto R+d_Hc$ the last term changes by
\begin{equation}
\exp\left[2\pi i \int_M (c_3\wedge j_7-c_1\wedge j_6)\right].
\end{equation}
This term is the usual WZ coupling of D-brane or RR current to a topologically
trivial RR field.

\subsubsection{RR fields of type IIA: Example 2}
In this section we give an example of a type IIA action which is
perhaps less familiar to the reader. Suppose that the
$10$-dimensional manifold $M$ is a product space $\Rh\times N$. The
most general metric on $M$ is
\begin{equation}
ds_M^2=-\rho^2 dt^2+(g_N)_{ij}(dx^i-\xi^i dt)(dx^j-\xi^j dt).
\label{lsmetric}
\end{equation}
where   $\rho$ is the  lapse and $\xi^j$ the
shift and $g_N$ is a Riemannian metric on $N$. From
the geometric point of view $\Theta = (dx^i-\xi^i dt)
\otimes \frac{\pd}{\pd x^i}$ is a connection on a topologically trivial bundle
$N\to M\to \Rh$
(for more details see section~7.2.2
of \cite{Belov:2006jd}).
The vector field $\pd/\pd t$ on $\Rh$ lifts to a vector field $\xi_M$ on $M$:
\begin{equation}
i_{\xi_M}\Theta=0\quad\Rightarrow\quad \xi_M:=\frac{\pd}{\pd t}+\xi^i\frac{\pd}{\pd x^i}.
\end{equation}
$\xi_M$ is a tangent vector defined by the proper time.
The connection $\Theta$ defines a decomposition of the tangent plane $T_xM$
into horizontal and vertical vectors. This decomposition is the orthogonal
decomposition in the metric \eqref{lsmetric}.

An orthogonal projector onto the space of horizontal vectors is
defined by
\begin{equation*}
P(\eta):=\xi_M\,\frac{g(\xi_M,\eta)}{g(\xi_M,\xi_M)}=\xi_M\, dt(\eta)
\quad\text{for}\quad
\eta\in \mathrm{Vect}(M).
\end{equation*}
Its dual $P^*:=dt\wedge  i_{\xi_M}$ defines a decomposition
of the differential forms into vertical and horizontal.

The  projector $P^*$   decomposes the space of even forms into
\begin{equation}
\Omega(M;\Rc)^0=
\underbrace{(1-P^*)\Omega(M;\Rc)^0}_{\text{vertical: }I(V_2)}
\oplus\underbrace{P^*\Omega(M;\Rc)^0}_{\text{horizontal: } V_2}
\label{HV}
\end{equation}

If in addition we assume that the $B$-field on $M$ is pulled back
from $N$ then we can choose our Lagrangian subspace $V_2$ to be the
horizontal forms and thus $I(V_2)$ is the space of vertical ones.
Since $M$ is a product space there is another decomposition of the
space of even forms
\begin{equation}
\Omega(M;\Rc)^0=\Omega^0(\Rh)\otimes \Omega(N;\Rc)^{0}
\oplus
\Omega^1(\Rh)\otimes \Omega(N;\Rc)^{-1}.
\label{rNdec}
\end{equation}

These two decomposition are related in the following way
\begin{equation}
R=\bar{R}_{\textsl{ev}}+dt\wedge \bar{R}_{\textsl{odd}}=
\underbrace{(\bar{R}_{\textsl{ev}}-dt\wedge i_{\xi}\bar{R}_{\textsl{ev}})}_{(R)_2^{\perp}}
+\underbrace{dt\wedge(\bar{R}_{\textsl{odd}}+i_{\xi}\bar{R}_{\textsl{ev}})}_{(R)_2}
\label{hvdec}
\end{equation}
where $\bar{R}_{\textsl{ev}}=\bar{R}_0+u^{-1}\bar{R}_2+\dots+
u^{-4}\bar{R}_8$,
$\bar{R}_{\textsl{odd}}=u^{-1}\bar{R}_1+u^{-2}\bar{R}_3+\dots+
u^{-5}\bar{R}_9$ and $\{\bar{R}_p\}$ are $t$-dependent
$p$-forms on $N$. Substituting the decomposition \eqref{hvdec} into
\eqref{Maction} one finds the action
\begin{multline}
S_L(R)=-\pi\int_{\Rh\times N}\rho\,dt\wedge\bigl[
\ell_s^{-10}\,\bar{R}_0\wedge *_N\bar{R}_0+\dots
+\ell_s^6\,\bar{R}_8\wedge *_N\bar{R}_8
\bigr]
\\
+\pi\int_{\Rh\times N}dt\wedge\bigl[
(\bar{R}_1+i_{\xi}\bar{R}_2)\wedge\bar{R}_8
\mp\dots-(\bar{R}_7+i_{\xi}\bar R_8)\wedge \bar R_2
+\bar{R}_9\wedge \bar{R}_0
\bigr]
\end{multline}
and the self-dual field
\begin{multline}
\mathcal{F}^+(R)=
\bar{R}_0+u^{-1}\bigl(\bar{R}_2-\ell_s^6\,\rho dt\wedge *_N\bar{R}_8\bigr)
+u^{-2}(\bar{R}_4+\ell_s^2\,\rho dt\wedge *_N\bar{R}_6)
\\
+u^{-3}(\bar{R}_6-\ell_s^{-2}\,\rho dt\wedge *_N\bar{R}_4)
+u^{-4}(\bar{R}_8+\ell_s^{-6}\,\rho dt\wedge *_N\bar{R}_2)
-u^{-5}\,\ell_s^{-10}\,\rho dt\wedge *_N\bar{R}_0.
\label{SDex2}
\end{multline}

Let us compare expressions for the self-dual fields
\eqref{SDex1} for product space and \eqref{SDex2}: it is clear that they are just
different parameterizations of the \textit{same} classical object $\mathcal{F}^+$.

\subsubsection{RR fields of type IIB: Example 1}
To construct an action principal we have to choose a Lagrangian
subspace. So consider a product space $M=\Rh\times N$
equipped with metric \eqref{lsmetric}.
The space of odd forms decomposes as
\begin{equation}
\Omega(M;\Rc)^{-1}
=\underbrace{u^{-1}\Omega^{1}\oplus u^{-2}\Omega^3\oplus
u^{-3}(1-P^*)\Omega^5}_{I(V_2)}\oplus
\underbrace{u^{-3}P^*\Omega^5\oplus u^{-4}\Omega^7\oplus
u^{-5}\Omega^9}_{V_2}
\end{equation}
where $P^*$ is the projection operator defined in the previous section.
If in addition we assume that the $B$-field on $M$ is pullback from
$N$ then we can choose Lagrangian subspace $V_2$ as shown above.

So any element $R$ of $V_1$ can be written as
\begin{equation}
R=\underbrace{u^{-1}R_1+u^{-2}R_3+u^{-3}(\bar{R}_5-dt\wedge i_{\xi}\bar{R}_5)}_{(R)_2^{\perp}}
+\underbrace{u^{-3}dt\wedge (\bar{R}_4+i_{\xi}\bar{R}_5)
+u^{-4}R_7+u^{-5}R_9}_{(R)_2}.
\label{RBex1}
\end{equation}
Substituting the decomposition \eqref{RBex1} into
\eqref{Maction} one finds the action
\begin{multline}
e^{iS_L(R)}=
\exp\left[-i\pi\int_{\Rh\times N}
\bigl[\ell_s^{-8} R_1\wedge *_M R_1+\ell_s^{-4}R_3\wedge *R_3
\bigr]
-i\pi\int_{\Rh\times N}\rho\,dt\wedge \bar{R}_5\wedge *_N\bar{R}_5
\right.
\\
\left.+i\pi\int_{\Rh\times N}\bigl[
dt\wedge(\bar{R}_4+i_{\xi}\bar{R}_5)\wedge \bar{R}_5
-R_7\wedge R_3+R_9\wedge R_1
\bigr]
\right].
\label{SIIBex1}
\end{multline}
and the self-dual field
\begin{equation}
\mathcal{F}^+(R)=
u^{-1}R_1+u^{-2}R_3+u^{-3}(\bar{R}_5-\rho dt\wedge *_N\bar{R}_5)
+u^{-4}\,\ell_s^{-4}*_M R_3-u^{-5}\,\ell_s^{-8}*_M R_1.
\label{SDBex1}
\end{equation}

\paragraph{Comparing the action \eqref{SIIBex1} to the  pseudo action.}
The equations of motion for IIB are usually obtained from a
``pseudo-action.'' The adjective  ``pseudo'' refers to the property
that its variation does not give the proper equations of motion.
Instead, to obtain the type IIB equations of motion one has to
impose the self-duality constraint by hand. The pseudo-action
contains the following Chern-Simons term\footnote{We are talking
here about the pseudo action defined in \cite{Polchinski:1998rr}.
There is a different pseudo action defined in
\cite{Bergshoeff:2001pv}, and it does not have this problem.}
\begin{equation}
\Phi_B=\exp\Bigl[-i\pi\int_{Y}\tilde{R}_5\wedge\tilde{H}\wedge  \tilde{R}_3\Bigr]
\label{PhiB}
\end{equation}
where $Y$ is an $11$-manifold bounding the $10$-manifold $M$. The
right hand side does not depend on a choice of extension provided
the integral of $\tilde{H}\wedge\tilde{R}_5\wedge  \tilde{R}_3$ over
any closed $11$-manifold is an even integer. In general, this is in
fact not the case! (Indeed this  can create difficulties in the
AdS/CFT correspondence \cite{Belov:2004ht}.) Let us compare this
Chern-Simons term with the one contained in the action
\eqref{SIIBex1}.

Suppose that $M$ is a product space $S^1\times N$ where $N$ is a
compact $9$-manifold. We choose our  $11$-manifold $Y$ to be
$D_2\times N$ where $D_2$ is a two dimensional disk bounding $S^1$.
Assuming that the RR fields have an extension to   $Y$ we can
rewrite the second term in \eqref{SIIBex1} as
\begin{equation*}
e^{+i\pi\int_{\pd D_2\times N}[
dt\wedge(\bar{R}_4+i_{\xi}\bar{R}_5)\wedge \bar{R}_5
-R_7\wedge R_3+R_9\wedge R_1
]}\stackrel{\text{Bianchi id.}}{=}\exp\left[
-2\pi i\int_{D_2\times N} \bar{R}_5\tilde{H}\tilde{R}_{3}
\right].
\end{equation*}
Notice the factor of $2$ as compared to \eqref{PhiB}!

\subsubsection{RR fields of type IIB: Example 2}
In this example we choose a different Lagrangian decomposition of
the space of odd forms on the product space $M=\Rh\times N$ equipped
with metric \eqref{lsmetric}. The space of odd forms now decomposes
as
\begin{equation}
\Omega(M;\Rc)^{-1}
=\underbrace{(1-P^*)\Omega(M;\Rc)^{-1}}_{I(V_2)}\oplus
\underbrace{P^*\Omega(M;\Rc)^{-1}}_{V_2}
\end{equation}
where $P^*$ is the projection operator defined in the previous
section. If in addition we assume that the $B$-field on $M$ is
pulled back from $N$ then we can choose Lagrangian subspace $V_2$ as
shown above. Since $M$ is a product space there is another
decomposition of the space of odd forms
\begin{equation}
\Omega(M;\Rc)^{-1}=\Omega^0(\Rh)\otimes \Omega(N;\Rc)^{-1}
\oplus
\Omega^1(\Rh)\otimes \Omega(N;\Rc)^{-2}.
\label{dec2}
\end{equation}

These two decomposition are related in the following way
\begin{equation}
R=\bar{R}_{\textsl{odd}}+dt\wedge \bar{R}_{\textsl{ev}}=
\underbrace{(\bar{R}_{\textsl{odd}}-dt\wedge i_{\xi}\bar{R}_{\textsl{odd}})}_{(R)_2^{\perp}}
+\underbrace{dt\wedge(\bar{R}_{\textsl{ev}}+i_{\xi}\bar{R}_{\textsl{odd}})}_{(R)_2}
\label{hvdec2}
\end{equation}
where $\bar{R}_{\textsl{odd}}=u^{-1}\bar{R}_1+\dots+
u^{-5}\bar{R}_9$,
$\bar{R}_{\textsl{ev}}=u^{-1}\bar{R}_0+\dots+
u^{-5}\bar{R}_8$ and $\{\bar{R}_p\}$ are $t$-dependent
$p$-forms on $N$. Substituting the decomposition \eqref{hvdec2} into
\eqref{Maction} one finds the action
\begin{multline}
S_L(R)=-\pi\int_{\Rh\times N}\rho\,dt\wedge\bigl[
\ell_s^{-8}\,\bar{R}_1\wedge *_N\bar{R}_1+\dots
+\ell_s^8\,\bar{R}_9\wedge *_N\bar{R}_9
\bigr]
\\
+\pi\int_{\Rh\times N}dt\wedge\bigl[
(\bar{R}_0+i_{\xi}\bar{R}_1)\wedge\bar{R}_9
-(\bar{R}_2+i_{\xi}\bar{R}_3)\wedge\bar{R}_7
\mp\dots
+(\bar{R}_8+i_{\xi}\bar{R}_9)\wedge \bar{R}_1
\bigr]
\label{SIIBex2}
\end{multline}
and the self-dual field
\begin{multline}
\mathcal{F}^+(R)=
u^{-1}\bigl(\bar{R}_1-\ell_s^{8}\,\rho dt\wedge *_N\bar{R}_9\bigr)
+u^{-2}(\bar{R}_3+\ell_s^{4}\,\rho dt\wedge *_N\bar{R}_7)
\\
+u^{-3}(\bar{R}_5-\rho dt\wedge *_N\bar{R}_5)
+u^{-4}(\bar{R}_7+\ell_s^{-4}\,\rho dt\wedge *_N\bar{R}_3)
+u^{-5}(\bar{R}_9-\ell_s^{-8}\,\rho dt\wedge *_N\bar{R}_1).
\label{SDBex2}
\end{multline}

\section{Dependence on metric}
\label{sec:metric}
\setcounter{equation}{0}
To study metric dependence of the partition function and
the action one has to use the following two results:
\begin{lemma}
Let $(M,g)$ be a $10$-dimensional Lorentzian manifold. The variation
of the involution $I$ defined in \eqref{Idef} with respect to the
metric is given by
\begin{equation}
\delta_gI=-\frac{1}{2}\,\tr (\xi_g)\, I+I\circ \xi_g
\quad\text{where}\quad
\xi_g=(\delta g^{-1}g)^{\mu}{}_{\nu}\,dx^{\nu}\wedge i(\tfrac{\pd}{\pd x^{\mu}}).
\label{xig}
\end{equation}
\end{lemma}

\begin{lemma}
If the Lagrangian subspace $V_2$ is chosen to be independent of the
 metric then the variation of the bilinear form
$\omega_j(v,\mathcal{F}^+(w))$ with respect to metric is a symmetric
form given by
\begin{equation}
\delta_g\omega_j(v,\mathcal{F}^+(w))
=\frac12\,\omega_j\bigl(\mathcal{F}^+(v),\,I\circ\xi_g\mathcal{F}^+(w)\bigr)
=-\frac12\,g_j\bigl(\mathcal{F}^+(v),\,\xi_g\mathcal{F}^+(w)\bigr).
\end{equation}
\end{lemma}

These two lemmas easily follow from
the results of section~8 in \cite{Belov:2006jd}.

\paragraph{Stress-energy tensor for the self-dual field.}
In section~\ref{sec:examples} we derived the following action for the
RR field on a Lorentzian manifold $(M,g)$:
\begin{equation}
S_L(R)=\pi\omega_j(R,\mathcal{F}^+(R))
\label{Mact}
\end{equation}
where $R$ is a $d_H$-closed form belonging to the isotropic subspace $\bar{V}_1$.
The stress-energy tensor is given by the following theorem
\begin{thm}
The variation of the action \eqref{Mact} with respect to the metric
is
\begin{equation}
\delta_g S_L(R)=\frac{\pi}{2}\,\omega_j\bigl(\mathcal{F}^+(R),\,I\circ\xi_g\mathcal{F}^+(R)\bigr)
=:-\frac{\pi}{2}\int_M\delta g^{\mu\nu}T_{\mu\nu}(\mathcal{F}^{+})\vol(g)
\label{dmetricS}
\end{equation}
where $\mathcal{F}^+(R):=(R)_2^{\perp}+I(R)_2^{\perp}$
is the self-dual projection of $R$, and the  operator
$\xi_g$ is defined in \eqref{xig}. The derivation of \eqref{dmetricS} relies only on
the fact that the subspaces $V_2$ and $\bar{V}_1$
do not depend on a choice of metric.
\end{thm}

In \cite{Belov:2006jd} we derived an expression for the stress-energy tensor of a $p$-form
using the operator $\xi_g$.
It is convenient to rewrite \eqref{dmetricS}
by substituting $\mathcal{F}^+(R)=(R)_2^{\perp}+I(R)_2^{\perp}$:
\begin{equation}
\int_M\delta g^{\mu\nu}T_{\mu\nu}\,\vol(g)=
g_j\bigl(\mathcal{F}^+(R),\xi_g(R)_2^{\perp}\bigr)
-\frac12\,\tr\xi_g\,g_j\bigl((R)_2^{\perp},(R)_2^{\perp}\bigr).
\end{equation}
where $\tr\xi_g = \tr(\delta g^{-1} g)$.

 Let us consider several examples. The self-dual field
$\mathcal{F}^+$ can be parameterized in many ways depending on the
choice of Lagrangian subspace $V_2$. Here we will consider the
self-dual fields presented in \eqref{F+IIA} and \eqref{F+IIB}.

\vspace{0.5cm}
\example{For type IIA one finds
\begin{multline}
\delta g^{\mu\nu}T_{\mu\nu}^{\text{IIA}}\,\vol(g)
=\ell_s^{-10}\,\Bigl[R_0\wedge *\xi_gR_0
-\frac12 \tr \xi_g\,R_0\wedge *R_0\Bigr]
\\
+\ell_s^{-6}\,\Bigl[R_2\wedge *\xi_gR_2
-\frac12 \tr \xi_g\,R_2\wedge *R_2\Bigr]
+\ell_s^{-2}\,\Bigl[R_4\wedge *\xi_gR_4
-\frac12 \tr \xi_g\,R_4\wedge *R_4\Bigr].
\label{TIIA}
\end{multline}
This is the standard stress-energy tensor for RR fields of type IIA.
}

\example{
For type IIB one finds
\begin{multline}
\delta g^{\mu\nu}T_{\mu\nu}^{\text{IIB}}\,\vol(g)
=\ell_s^{-8}\,\Bigl[R_1\wedge *\xi_gR_1
-\frac12 \tr \xi_g\,R_1\wedge *R_1\Bigr]
\\
+\ell_s^{-4}\,\Bigl[R_3\wedge *\xi_gR_3
-\frac12 \tr \xi_g\,R_3\wedge *R_3\Bigr]
+\mathcal{F}^+_5\wedge *\xi_g\mathcal{F}^+_5.
\label{TIIB}
\end{multline}
This is the standard stress-energy tensor for RR fields of type IIB.
}

\section{Conclusion: Future Directions}
\label{sec:conc}
\setcounter{equation}{0}

There are many potentially fruitful   directions for future
research.  As mentioned in the introduction, one of our main
motivations was to understand the action sufficiently clearly to be
able to address the problem of computing the amplitudes for brane
instanton mediated transitions between flux vacua. It is also of
interest to compute more explicitly the one-loop determinants,
especially in the IIB case (this is also true, and even more
pressing,  for the self-dual field on the M5-brane).  We intend to
use our improved understanding of the twisted K-theory partition
function to understand how the computation of
\cite{Diaconescu:2000wy} generalizes to the case of nontrivial
background $H$-flux. It would also be very interesting to see if our
formalism fits in well with supersymmetry, and in particular how the
fermions in supergravity can be incorporated. The results of
\cite{Bergshoeff:2001pv} make this avenue of research appear to be
very promising. On the more mathematical end, one of the weaknesses
in our discussion has been the treatment of differential cocycles.
Indeed our motivation for regarding a differential K-theory group as
a gauge group was somewhat formal. It would be very nice to have a
concrete model of differential cocycles in twisted differential
K-theory which makes this identification more obvious. Furthermore,
there are interesting formal similarities between the discussion of
twisted Chern characters and certain aspects of generalized complex
geometry which we hope to explore.

\section*{Acknowledgments}
We would like to thank D. ~Freed for initial collaboration on this
project, and J. ~Distler and D. ~Freed for collaboration on closely
related projects.  We would also like to thank S.~de~Alwis,
  J.~ Harvey, J.~Jenquin, S.~Lukic, S.~Reffert, G.~Segal, Y. Tachikawa,
   E.~Witten and L.~Pando
Zayas for discussions and correspondence. D.B. thanks C.~Hull, A.-K. Kashani-Poor,
K.~Stelle, A.~Tseytlin and D.~Waldram for discussions and
correspondence. We would like to acknowledge the hospitality of the
Kavli Institute for Theoretical Physics, the Aspen Center for
  Physics, and the Simons Workshop at Stony Brook,  where
part of this work was done. G.M. thanks the Institute for Advanced
Study and the Monell foundation for hospitality. D.B. thanks New
High Energy Theory Center at Rutgers University, the Center for
Theoretical Physics at University of Michigan and Theoretical
Physics Group at the University of California at Berkeley. This work
was supported in part by DOE grant DE-FG02-96ER40949, and by the
National Science Foundation under Grant No. PHY99-07949. D.B. was
supported by PPARC and in part by RFBR grant 05-01-00758.

\clearpage
\appendix
\renewcommand {\theequation}{\thesection.\arabic{equation}}

\section{Twisted cohomology}
\label{sec:appA}
\setcounter{equation}{0}

In this appendix we give an example of some nontrivial
twisted cohomology groups, following the work of
 Atiyah and Segal  \cite{AtiyahSegal}.

Twisted cohomology is a cohomology for the twisted differential
$d_H=d-u^{-1}H$ which acts on
$\Omega(X;\Rh[u,u^{-1}])$. It is clear that $d_H$ operator preserves only parity
and therefore there are only two twisted cohomologies: even $H^0_{d_H}(X)$ and
odd $H^1_{d_H}(X)$. Although the twisted differential does not preserve
grading of the de Rham complex it preserves filtration whose $p$-component
is the sum of the forms of degrees grater or equal than $p$.
Atiyah and Segal \cite{AtiyahSegal} showed that this filtration yields
a spectral sequence which converges to the twisted cohomology.

The lowest term in the spectral sequence is just $\Omega^{\bullet}(X;\Rh)$
with the usual differential $d$, so
\begin{equation*}
E_2=H^{\bullet}(X).
\end{equation*}
Now consider $x$ of the form
\begin{equation}
x=x_p+u^{-1}x_{p+2}+u^{-2}x_{p+4}+\dots
\end{equation}
where $dx_p=0$ is a closed $p$-form representing the cohomology class $[x_p]\in H^p(X)$.
We want to choose $x_{p+2},\,x_{p+4},\dots$ such that $d_H x=0$. This yields the following
series of equations
\begin{equation}
dx_{p+2}-H\wedge x_p=0,\quad d x_{p+4}-H\wedge x_{p+2}=0,\quad \text{etc}.
\label{serofeqs}
\end{equation}
The first equation says that $H\wedge x_p$ vanishes in the de Rham cohomology.
Thus the third differential in the spectral sequence $d_3$ is just multiplication
by $-H$. Hence $E_3=\ker d_3/\mathrm{im}\, d_3$.
There is no differential of degree $4$ so $E_4\cong E_3$. Now for $x_p\in E_4$
it guarantied that $[H\wedge x_p]=0$ thus there exists $x_{p+2}$ such
that $dx_{p+2}=H\wedge x_p$. The second equation in \eqref{serofeqs} says
that $H\wedge x_{p+2}$ must vanish in the de Rham cohomology. This imposes
a restriction on $x_p$ which can be written using Massey product
$\{H,H,x_p\}=0$ (for definition of the Massey product see \cite{AtiyahSegal}).
Thus the fifth differential in the spectral sequence is $d_5=-\{H,H,\cdot\}$.
So $E_6\cong E_5=\ker d_5/\mathrm{im}\,d_5$. In \cite{AtiyahSegal} it is proved
that the higher differentials of this spectral sequence are given by the higher
Massey products, e.g. $d_7=-\{H,H,H,\cdot\}$ etc.
Thus a $d_H$-cohomology is roughly speaking a subspace of all cohomology classes
which are annihilated by the higher differentials (modulo images). Given an element $x_p$ from
this subspace one can construct $x_{p+2},\,x_{p+4},\dots $ such
that $x=x_p+u^{-1}x_{p+2}+\dots$ is $d_H$-closed. If $x_p$ is not in the image
of higher differentials then $x$ represents a nontrivial $d_H$-cohomology class.

\paragraph{Examples.}
Now we consider an example.
In Appendix A of \cite{AtiyahSegal} it was constructed a simple
$n+1$-manifold $Y_n$ which has nonvanishing Massey products, and
thus interesting $d_H$ cohomology.

\paragraph{$Y_2$.}
$Y_2$ is a twisted circle bundle over $S^1\times S^1$. Let $x,y$ be closed
one forms with integral periods representing generators $[x]$ and $[y]$
of $H^1(S^1\times S^1)$. We denote by $z$ a connection one form with
$dz=xy$. The de Rham cohomology of $Y_2$ has the following generators
\begin{alignat*}{3}
1 \quad & \text{in}\quad & H^0(Y_2);
\\
[x],\,[y]\quad & \text{in}\quad & H^1(Y_2);
\\
[xz],\,[yz]\quad & \text{in}\quad & H^2(Y_2);
\\
[xyz]\quad & \text{in}\quad & H^3(Y_2).
\end{alignat*}
The nonvanishing Massey products are $\{x,x,y\}=xz$ and $\{y,y,x\}=-yz$.

To obtain a $3$-form we consider the $7$-manifold $Y_2\times \Ch P^2$.
We denote by $t$ a closed $2$-form with integral periods which
represents a generator $[t]$ of $H^*(\Ch P^2)$.
We choose $H=xt$. The nonzero differentials are $d_3\phi=-xt\phi$
and $d_5\phi=-\{xt,xt,\phi\}$ ($d_5y=xzt^2$).
A simple calculation shows that the ranks of
$E_2, E_4$ and $E_6=E_{\infty}$ are $18,\,10$ and $8$.
$H_{d_H}(Y_2\times \Ch P^2)$ is generated by
\begin{equation}
\begin{tabular}{|l||c|c|c||c|c|c|}
\hline
\rowcolor[gray]{0.85}
$p$ & $1$ & $3$ & $5$ & $2$  & $4$  & $6$
\\
\hline
$x_p$ & $x$ & $yt,\,xyz$ & $yt^2$ & $xz$  & $xzt,\,t^2$  & $yz t^2$
\\
\hline
\end{tabular}
\end{equation}
To get a $d_H$-cohomology class one has to restore $x_{p+2},$ etc, e.g.
\begin{equation*}
[x]_{d_H}=x,\quad [xz]_{d_H}=xz,\quad [yt]_{d_H}=yt +u^{-1}zt^2,\quad\text{etc}.
\end{equation*}
As it is expected the dimension of the twisted cohomology, say $\dim H^0_{d_H}=4$,
is smaller then the sum of the even Betti numbers $=9$.

\paragraph{$Y_3$.} $Y_3$ is a twisted circle bundle over $Y_2$.
If we denote by $v$ a connection $1$-form the its curvature
is one of the previous Massey products, say $dv=xz$.
The de Rham cohomology of $Y_3$ has the following generators
\begin{alignat*}{3}
1 \quad & \text{in}\quad & H^0(Y_3);
\\
[x],\,[y]\quad & \text{in}\quad & H^1(Y_3);
\\
[xv],\,[yz]\quad & \text{in}\quad & H^2(Y_3);
\\
[yzv],\,[xzv]\quad & \text{in}\quad & H^3(Y_3);
\\
[xyzv]\quad & \text{in}\quad & H^4(Y_3).
\end{alignat*}
The nonvanishing Massey products are $\{x,x,x,y\}=xv$ and $\{y,y,x\}=-yz$.

To obtain a $3$-form we consider a $10$-manifold $X=Y_3\times \Ch P^3$.
We denote by $t$ a closed $2$-form with integral periods which
represents a generator $[t]$ of $H^*(\Ch P^3)$.
We will consider two twistings $H=xt$ and $H=yt$.
The nonvanishing Massey products are $\{xt,xt,xt,y\}=xvt^3$,
$\{yt,yt,x\}=-yz t^2$ and $\{yt,yt,xt\}=-yz t^3$.
For $H=xt$ the ranks of $E_2,\,E_4=E_6$ and $E_8=E_{\infty}$ are
$32,\,20$ and $16$. The $d_{xt}$-cohomologies $H^0_{d_{xt}}$ and $H^1_{d_{xt}}$
are generated by
\begin{equation*}
\begin{tabular}{|l||c|c|c|c||c|c|c|c|c|}
\hline
\rowcolor[gray]{0.85}
$p$ & $2$ & $4$ & $6$ & $8$ & $1$ & $3$ & $5$ & $7$ & $9$
\\
\hline
$x_p$ & $xv$ & $xyzv,\,xvt,\,yzt$ & $xvt^2,\,t^3,\,yzt^2$ & $yzt^3$
& $x$ & $xzv,\,yt$ & $yt^2,\,xzvt$ & $yt^3,\,xzvt^2$ & $yzvt^3$

\\
\hline
\end{tabular}
\end{equation*}
To get a $d_{xt}$-cohomology class one has to restore $x_{p+2},$ etc, e.g.
\begin{equation*}
[xv]_{d_{xt}}=xv,\quad [yzt]_{d_{xt}}=yzt+u^{-1} yvt^2+u^{-2}zv t^3,\quad
[yt]_{d_{xt}}=yt +u^{-1}zt^2+u^{-2}vt^3,\quad\text{etc}.
\end{equation*}
One sees that rank of $K^0_{xt}$ is $8$.

For $H=yt$ the ranks of $E_2,\,E_4$ and $E_6=E_{\infty}$ are
$32,\,20$ and $12$. The $d_{yt}$-cohomologies $H^0_{d_{yt}}$ and $H^1_{d_{yt}}$
are generated by
\begin{equation}
\begin{tabular}{|l||c|c|c|c||c|c|c|c|c|}
\hline
\rowcolor[gray]{0.85}
$p$ & $2$ & $4$ & $6$ & $8$ & $1$ & $3$ & $5$ & $7$ & $9$
\\
\hline
$x_p$ & $yz$ & $xyzv,\,yzt$ & $xvt^2,\,t^3$ & $xvt^3$
& $y$ & $yzv$ & $xt^2,\,yzut$ & $xt^3$ & $xzvt^3$
\\
\hline
\end{tabular}
\end{equation}
To get a $d_{yt}$-cohomology class one has to restore $x_{p+2},$ etc, e.g.
\begin{equation*}
[xvt^2]_{d_{yt}}=xvt^2-u^{-1} zvt^3,\quad
[xt^2]_{d_{yt}}=xt^2 -u^{-1}zt^3,\quad\text{etc}.
\end{equation*}
One sees that the rank of $K^0_{yt}$ is $6$.

\section{Determining the metric-dependent factor }
\label{sec:appC}
\setcounter{equation}{0}

In \cite{Belov:2004ht,Belov:2006jd} we found that normalizing the
Chern-Simons wavefunction $\|\mathcal{Z}\|^2=1$ identifies $\Nc_g$,
or more precisely, $\|\Nc_{V_1,V_2}(g)\|^2$,  with the correct
one-loop determinants of the holographically dual field. It is not
clear \textit{a priori} why this is the correct procedure --- although
it fits in very beautifully with the viewpoint that the partition
function in AdS/CFT should be regarded as a wavefunction ---  but
since it works in other cases we will carry out that procedure for
the example of the RR fields.

The partition function $\mathcal{Z}(a;\check{j})$ restricted
to $P_{\check{j}}\cong \Omega(X;\Rc)^j_{d_H}/\Omega(X;\Rc)^j_{d_H,\Zh}$ defines
an element of the finite dimensional Hilbert space $\Hc^{qu}$.
One can normalize the wave function, $\|\mathcal{Z}\|^2=1$, with respect to the inner product in
$\Hc^{qu}$ and in this way fix the norm square of $\Nc_{V_1,V_2}(g)$.

It is clear that $\Nc_{V_1,V_2}(g)$ does not depend on the source $\check{j}$. So
to simplify the calculation we put it zero, and assume that the
characteristic class $\mu=0$. In this case we can choose
the base point $\check{A}_{\bullet}=0$. This means that
$\sigma(\check{A}_{\bullet})=0$ and
$a=\check{A}-\check{A}_{\bullet}$ is a $d_H$-closed form. In Theorem~\ref{thm62}
we introduced a normalization factor $\Nc_{V_1,V_2}(g)$. This must be regarded
as a section of a Hermitian line bundle $\mathscr{L}$  over the space
of metrics. The norm on the Hilbert space $\Hc^{qu}$ is just the
$L^2$-norm on $\mathscr{L}\otimes \Lb$:
\begin{equation}
\|\mathcal{Z}\|^2_{L^2}:=\int_{\Omega(X;\Rc)^j_{d_H}/\Omega(X;\Rc)^j_{d_H,\Zh}}\Ds
a\, \|\mathcal{Z}(a)\|^2 \label{norm}
\end{equation}
where the second set of $\| \cdot \|^2$ denotes the norm on
$\mathscr{L}$.

 {}From Theorem~\ref{thm62} and Corollary~\ref{cor61} we
learn that the partition function restricted to the real slice
$a^-=(a^+)^*$ can be written as
\begin{equation}
\mathcal{Z}\left[\begin{smallmatrix}
\varepsilon_1
\\
\varepsilon_2
\end{smallmatrix}
\right](a)=\Nc_{V_1,V_2}(g)\,e^{-i\pi\omega(\varepsilon_2,\varepsilon_1)}
\sum_{R\in \bar{\Gamma}_1^h-\varepsilon_1}
e^{-\frac{\pi}{2}\,(H-B)(R^+-a_1^+,R^+-a_1^+) +2\pi i
\omega(a_2+\varepsilon_2,R) +i\pi\omega(a_1,a_2)}
\end{equation}
where $a$ is \textit{a harmonic} form, $a=a_1+a_2$ according to the Lagrangian
decomposition $\Gamma^h=\Gamma_1^h\oplus\Gamma_2^h$.

To calculate the norm \eqref{norm} we need to fix a gauge in this
functional integral. This can be done by using equation
\eqref{nrmIIA}. By evaluating the Gaussian integral and solving
the equation $\|\mathcal{Z}\|^2_{L^2}=1$ for ${\Nc}_g$ one finds
for IIA:
\begin{subequations}
\begin{equation}
\|\mathcal{ N}_g\|^{2} =
\left[\det(\Im
\tau)\prod_{p=1}^{10}\left[
\frac{\mathrm{Vol}(\mathbb{T}^{-p})^{-2}\det{}'(d^*_{-p}d_{-p})^{1/2}}{
\mathrm{Vol}(\mathbb{T}^{-p,\,\bullet\geqslant 12-p})^{-2}
\det{}'(d^*_{-p}d_{-p}|_{\Omega^{\bullet\geqslant 12-p}})^{1/2}}
\right]^{(-1)^{p+1}}\right]^{1/2} \label{NA}.
\end{equation}
for IIB:
\begin{equation}
\|\mathcal{ N}_g\|^{2} =
\left[\det(\Im
\tau)\prod_{p=1}^{10}\left[
\frac{\mathrm{Vol}(\mathbb{T}^{-1-p})^{-2}\det{}'(d^*_{-1-p}d_{-1-p})^{1/2}}{
\mathrm{Vol}(\mathbb{T}^{-1-p,\,\bullet\geqslant 11-p})^{-2}
\det{}'(d^*_{-1-p}d_{-1-p}|_{\Omega^{\bullet\geqslant 11-p}})^{1/2}}
\right]^{(-1)^{p+1}}\right]^{1/2} \label{NB}.
\end{equation}
\label{N}
\end{subequations}

{}From equation \eqref{N} it follows that $\Nc_{V_1,V_2}(g)$ is some kind of
square root of the right hand side of \eqref{N}. We now conjecture
that there is a very natural squareroot provided we view $\Nc_{V_1,V_2}(g)$ as
a section of some determinant line bundle. We expect that we should
set
\begin{equation}
\eqref{NA}\quad\text{or}\quad\eqref{NB}=\|\det \overline{\mathcal{D}}\|^2_Q
\label{Quillen}
\end{equation}
where the right hand side is the Quillen norm of a section $\det
\overline{\mathcal{D}}$ of some determinant line bundle
$\texttt{DET}(\overline{\mathcal{D}})$ over the space of metrics on
$X$.

\paragraph{Gauge fields.} Consider a gauge potential $a\in\Omega(X;\Rc)^j$.
Denote by $\mathscr{G}^{j}$ the group of gauge transformations
$a\mapsto a+\omega^j$ where $\omega^j\in\Omega(X;\Rc)^j_{d_H,\Zh}$.
In this paragraph we want to obtain a formula for
\begin{equation*}
\int_{\Omega(X;\Rc)^j/\mathscr{G}^j}\Ds a.
\end{equation*}
Note that $d_j:\Omega(X;\Rc)^j\to \Omega(X;\Rc)^{j+1}$ ($d_j\omega:=d_H\omega$)
is an elliptic operator.
Moreover the corresponding  Laplacian $D_j=d_j^*d_j+d_{j-1}d_{j-1}^*$
is a self-adjoint positive elliptic operator where
$d_j^*$ denotes the adjoint operator in the Riemannian metric \eqref{Metdef}.
Note that the metric \eqref{Metdef} is a dimensionless metric defined by
a Riemannian metric $g_E$ on $X$ and the string scale $\ell_s$.
Any form $a\in \Omega(X;\Rc)^j$
can be uniquely written in the form (Hodge decomposition)
\begin{equation*}
a=a^h+a^T+d_{j-1}\alpha_{j-1}^T
\end{equation*}
where $a^h\in \ker D_j=:\mathscr{H}^j$ (twisted harmonic forms),
$a^T\in\mathrm{im}\, d^*_j$ and
$\alpha_{j-1}^T\in\mathrm{im}\, d^*_{j-1}$. This implies
\begin{equation}
  \|\delta a\|^2_{j}=\|\delta a^h\|^2_{j}
+\|\delta a^T\|^2_{j} +g_{j-1}\bigl(\delta\alpha^T_{j-1},
[d^{*}_{j-1}d_{j-1}]\delta\alpha^T_{j-1}\bigr)
\end{equation}
where $\|w\|_j^2$ is $g_j(w,w)$.
 Thus
\begin{equation}
\int_{\Omega(X;\Rc)^j}\Ds a=\int_{\mathscr{H}^{j}}\Ds a^h\int_{\mathrm{im}\,d^*_j}\Ds a^T
\int_{\mathrm{im}\, d_{j-1}^*}\Ds\alpha_{j-1}^{T}\,
\bigl[\det{}'(d^{*}_{j-1}d_{j-1})\bigr]^{1/2}.
\label{Da}
\end{equation}
The gauge group $\mathscr{G}^j$ has several connected
components labelled by the harmonic forms
with quantized periods $\mathscr{H}^{j}_{\Zh}(X)$. Using the Hodge
decomposition we can write
\begin{equation}
\int_{\Omega(X;\Rc)^j/\mathscr{G}^j}\Ds a
=\int_{\mathscr{H}^j/\mathscr{H}^{j}_{\Zh}}\Ds a^h\int_{\mathrm{im}\,d^*_j}
\Ds a^T
\int_{\mathrm{im}\, d_{j-1}^*}
\frac{ \Ds\alpha_{j-1}^{T}}{\mathrm{Vol}(\mathscr{G}^{j}_0)}\,
\bigl[\det{}'(d^{*}_{j-1}d_{j-1})\bigr]^{1/2}
\label{Da2}
\end{equation}
where $\mathscr{G}^{j}_{0}
\cong\Omega(X;\Rc)^j_{d_H-\text{exact}}/\mathscr{G}^{j-1}$ is the connected
component of the identity of the gauge group $\mathscr{G}^{j}$.

\paragraph{Volume of the gauge group (motivating example).}
Before considering the general expression for the volume of the
gauge group we consider the example of type IIA: in this case our
gauge field is $a\in \Omega(X;\Rc)^0$ and the corresponding gauge
transformations and gauge transformations for gauge transformations
can be summarized by
\begin{align*}
\Omega(X;\Rc)^{-0}&=u^{-0}\Omega^0\oplus u^{-1}\Omega^2\oplus u^{-2}\Omega^4
\oplus u^{-3}\Omega^6\oplus u^{-4}\Omega^8 \oplus u^{-5}\Omega^{10};
\\
\Omega(X;\Rc)^{-1}&=\fcolorbox[gray]{0.6}{0.9}{$u^{-0}\Omega^1\oplus u^{-1}\Omega^3\oplus u^{-2}\Omega^5
\oplus u^{-3}\Omega^7\oplus u^{-4}\Omega^9$};
\\
\Omega(X;\Rc)^{-2}&=\fcolorbox[gray]{0.6}{0.9}{$u^{-1}\Omega^0\oplus u^{-2}\Omega^2
\oplus u^{-3}\Omega^4\oplus u^{-4}\Omega^6 \oplus u^{-5}\Omega^{8}$}
\oplus u^{-6}\Omega^{10};
\\
\Omega(X;\Rc)^{-3}&=\fcolorbox[gray]{0.6}{0.9}{$u^{-2}\Omega^1\oplus u^{-3}\Omega^3\oplus u^{-4}\Omega^5
\oplus u^{-5}\Omega^7$}\oplus u^{-6}\Omega^9;
\\
\Omega(X;\Rc)^{-4}&=\fcolorbox[gray]{0.6}{0.9}{$u^{-2}\Omega^0\oplus u^{-3}\Omega^2
\oplus u^{-4}\Omega^4\oplus u^{-5}\Omega^6$} \oplus u^{-6}\Omega^{8}
\oplus u^{-7}\Omega^{10};
\\
&\vdots
\\
\Omega(X;\Rc)^{-9}&=\fcolorbox[gray]{0.6}{0.9}{$u^{-5}\Omega^1$}
\oplus u^{-6}\Omega^3\oplus u^{-7}\Omega^5
\oplus u^{-8}\Omega^7\oplus u^{-9}\Omega^9;
\\
\Omega(X;\Rc)^{-10}&=\fcolorbox[gray]{0.6}{0.9}{$u^{-5}\Omega^0$}
\oplus u^{-6}\Omega^2
\oplus u^{-7}\Omega^4\oplus u^{-8}\Omega^6 \oplus u^{-9}\Omega^{8}
\oplus u^{-10}\Omega^{10}.
\end{align*}
The forms in the gray boxes correspond to the ``real'' gauge transformations
while the forms outside the gray boxes correspond to the fields which has to be excluded.
Recall that the group defined on each line acts on the previous  one by
$\alpha_{-k}\mapsto \alpha_{-k}+d_H\alpha_{-k-1}$. If the curvature $H$ of the $B$-field
were zero when we could easily describe the gauge group appearing on each line
--- it consists just of the elements in the gray box modulo closed forms
with quantized periods. But if $H\ne 0$ the description is slightly
more complicated because $d_H$ does not preserve the space indicated
by the gray boxes. We must quotient out by the terms not appearing
in the gray boxes.  The gauge group appearing on each line is
\begin{equation}
\mathcal{F}^{-k}:=\Bigl[\Omega(X;\Rc)^{-k}/\Omega(X;\Rc)^{-k}_{d_H,\Zh}\Bigr]
/\Bigl[\Omega(X;\Rc)^{-k,\,\bullet\geqslant 12-k}/\Omega(X;\Rc)^{-k,\,\bullet\geqslant 12-k}_{d_H,\Zh}
\Bigr]
\label{gaugegroupdef}
\end{equation}
where $\Omega(X;\Rc)^{-k,\,\bullet\geqslant 12-k}$ denotes the space of
forms of total degree $-k$ and differential form degree greater or equal $12-k$.
It is easy to see that for $H=0$ equation \eqref{gaugegroupdef} yields
the correct gauge group.

\paragraph{Volume of the gauge group.}
To calculate the volume $\mathrm{Vol}(\mathscr{G}^{j}_0)$ we
notice that
\begin{equation}
\mathrm{Vol}(\mathscr{G}^{j}_0)=
\int_{\Omega(X;\Rc)^j_{d_H-\text{exact}}/\mathscr{G}^{j-1}}\Ds \alpha_{j}=
\int_{\mathcal{F}^{j-1}} \Ds \alpha_{j-1}
=\frac{\int_{\Omega(X;\Rc)^{j-1}/\Omega(X;\Rc)^{j-1}_{d_H,\Zh}}\Ds\alpha_{j-1}}{
\int_{\Omega(X;\Rc)^{j-1\,\bullet\geqslant 11+j}/\Omega(X;\Rc)^{j-1\,\bullet\geqslant 11+j}_{d_H,\Zh}}\Ds\alpha_{j-1}}
\label{int1}
\end{equation}
where $\mathcal{F}^{j-1}$ is defined in \eqref{gaugegroupdef}. Note
that $j=0,-1$ are the two cases of interest. Effectively  we have
two different integrals: one in the numerator and another in the
denominator. The gauge groups the integrals in the numerator are
always $\Omega(X;\Rc)^{-k}_{d_H,\Zh}$ while the gauge groups for the
integrals in the denominator are always
$\Omega(X;\Rc)^{-k,\,\bullet\geqslant 12-k}_{d_H,\Zh}$.

Each of these integrals can now be calculated in the standard way (see, for example,
appendix~C in \cite{Belov:2006jd}). The final result is
\begin{equation}
\mathrm{Vol}(\mathcal{G}^j_0)=
\int_{\mathrm{im}\, d_{j-1}^*}\hspace{-5mm}\Ds\alpha_{j-1}^T\;
\prod_{p=1}^{10}
\left[\frac{\mathrm{Vol}(\mathbb{T}^{j-p})\det{}'(d^*_{j-p-1}d_{j-p-1})^{1/2}}{
\mathrm{Vol}(\mathbb{T}^{j-p,\,\bullet\geqslant 12+j-p})
\det{}'(d^*_{j-p-1}d_{j-p-1}|_{\Omega^{\bullet\geqslant 11+j-p}})^{1/2}}\right]^{(-1)^{p+1}}
\end{equation}
where
\begin{equation}
\mathbb{T}^{j-p}=\mathscr{H}^{j-p}/\mathscr{H}^{j-p}_{\Zh}
\quad\text{and}\quad
\mathbb{T}^{j-p,\,\bullet\geqslant 12+j-p}
=\mathscr{H}^{j-p,\,\bullet\geqslant 12+j-p}/\mathscr{H}^{j-1,\,\bullet\geqslant 12+j-p}_{\Zh}.
\end{equation}
Here $\mathscr{H}^{-k}$ is the space of twisted harmonic forms of total degree $-k$,
$\mathscr{H}^{-k}_{\Zh}$ is the space of twisted harmonic forms of total degree $-k$
with quantization condition and $\mathscr{H}^{-k,\,\bullet\geqslant 12-k}$
is the space of twisted harmonic forms of total degree $-k$ and differential
form degree $\geqslant 12-k$.
The volume of the twisted harmonic torus is
\begin{equation}
\mathrm{Vol}(\mathbb{T}^{-k})=\bigl[\det g_{-k}(\omega_{\alpha},\omega_{\beta})\bigr]^{1/2}
\label{C:Vp}
\end{equation}
$\{\omega_{\alpha}\}$ is an ``integral'' basis of twisted harmonic
forms of total degree $-k$. Notice that
$\mathrm{Vol}(\mathbb{T}^{-k})$ does not depend on the choice of the
``integral'' basis $\{\omega_{\alpha}\}$.

Combining this result with \eqref{Da2} one finds
\begin{subequations}
\begin{multline}
\int_{\Omega(X;\Rc)^0/\Omega(X;\Rc)^0_{d_H,\Zh}}\Ds a
=\int_{\mathscr{H}^0/\mathscr{H}^0_{\Zh}}\Ds a^h
\int_{\mathrm{im}\,d^*_0}\Ds a^T
\\
\times\left\{\prod_{p=1}^{10}\left[
\frac{\mathrm{Vol}(\mathbb{T}^{-p})^{-2}\det{}'(d^*_{-p}d_{-p})^{1/2}}{
\mathrm{Vol}(\mathbb{T}^{-p,\,\bullet\geqslant 12-p})^{-2}
\det{}'(d^*_{-p}d_{-p}|_{\Omega^{\bullet\geqslant 12-p}})^{1/2}}
\right]^{(-1)^{p+1}}\right\}^{1/2}
;
\label{nrmIIA}
\end{multline}
and
\begin{multline}
\int_{\Omega(X;\Rc)^{-1}/\Omega(X;\Rc)^{-1}_{d_H,\Zh}}\Ds a
=\int_{\mathscr{H}^{-1}/\mathscr{H}^{-1}_{\Zh}}\Ds a^h
\int_{\mathrm{im}\,d^*_{-1}}\Ds a^T
\\
\times\left\{\prod_{p=1}^{10}\left[
\frac{\mathrm{Vol}(\mathbb{T}^{-1-p})^{-2}\det{}'(d^*_{-1-p}d_{-1-p})^{1/2}}{
\mathrm{Vol}(\mathbb{T}^{-1-p,\,\bullet\geqslant 11-p})^{-2}
\det{}'(d^*_{-1-p}d_{-1-p}|_{\Omega^{\bullet\geqslant 11-p}})^{1/2}}
\right]^{(-1)^{p+1}}\right\}^{1/2}.
\label{nrmIIB}
\end{multline}
\end{subequations}
In these equations we assumed that $\mathrm{Vol}(pt)=1$.

%


\end{document}